\documentclass[12pt,a4paper]{article}
\pdfoutput=1
\usepackage[utf8]{inputenc}
\usepackage[T1]{fontenc}
\usepackage[margin=2.0cm]{geometry}
\usepackage[pdfstartview={FitH},bookmarks=false,linktoc=page,colorlinks=true,linkcolor=blue,citecolor=blue,urlcolor=blue]{hyperref}
\usepackage[numbered]{bookmark}
\usepackage{mathtools}
\usepackage{amssymb}
\usepackage{graphicx}
\usepackage[font=small,labelfont=bf]{caption}
\usepackage{authblk}
\usepackage{cite}
\usepackage{dsfont}
\usepackage{xcolor}
\usepackage[titles]{tocloft}
\usepackage{float}
\usepackage{subcaption}
\usepackage{color}
\usepackage{tikz}
\usepackage{ifthen}
\usepackage[warn]{textcomp}%for the number sign
\numberwithin{equation}{section}
\allowdisplaybreaks
\usepackage{multirow}

\usetikzlibrary{patterns}

\usepackage{comment}

\usepackage{pgfplots}
\pgfplotsset{compat=1.9}

\usetikzlibrary{arrows.meta, patterns, intersections, backgrounds}
\usepgfplotslibrary{fillbetween}

\tikzset{ntp/.style={circle, thin, minimum size=2mm, inner sep=0,
fill=white,#1}}

\usepackage{caption}
\usepackage{subcaption}

\begin{document}

%%%%%%%%%%%%%%%%%%%%%%%%%%%%%%%%%%%%%%%%%%%%
\title{\vspace{2cm}\textbf{Global and local thermodynamics of the (2$+$1)-dimensional rotating Gauss–Bonnet black hole}\vspace{1cm}}
%%%%%%%%%%%%%%%%%%%%%%%%%%%%%%%%%%%%%%%%%%%%

\author[a,b]{H. Dimov}
\author[a]{M. Radomirov}
\author[a]{I. N. Iliev}
\author[a,c]{R. C. Rashkov}
\author[a,b]{T. Vetsov}

\affil[a]{\textit{Department of Physics, Sofia University,}\authorcr\textit{5 J. Bourchier Blvd., 1164 Sofia, Bulgaria}
	
	\vspace{-10pt}\texttt{}\vspace{0.0cm}}

\affil[b]{\textit{The Bogoliubov Laboratory of Theoretical Physics, JINR,}\authorcr\textit{141980 Dubna,
		Moscow region, Russia}
	
	\vspace{-10pt}\texttt{}\vspace{0.0cm}}

\affil[c]{\textit{Institute for Theoretical Physics, Vienna University of Technology,}
  \authorcr\textit{Wiedner Hauptstr. 8–10, 1040 Vienna, Austria}
	\vspace{10pt}\texttt{h\_dimov,ivo.iliev,radomirov,rash,vetsov@phys.uni-sofia.bg}\vspace{0.1cm}}
\date{}
\maketitle

\begin{abstract}
{ The aim of this paper is to study the local and the global thermodynamic
properties of the 3-dimensional rotating Gauss-Bonnet black hole. To this
end we consider the conditions for local and global thermodynamic
stability of the solution in a given ensemble of state quantities.
Concerning the local analysis we found the regions of
stability for every physical specific heat together with the existing Davies
curves. Another central result is the generalization of the notion of global
thermodynamic stability, known from the standard thermodynamics, to describe the global
equilibrium of black holes. The new approach consists of applying specific Legendre transformation of
the energy or the entropy to find the natural thermodynamic potential for the
given ensemble of macro parameters. The global stability analysis, restricted to the 
week positivity conjecture is based on 
the properties of the new thermodynamic potential. The advantage of this method is that it allows one to chose different potentials,
corresponding to different constraints to which the system may be subjected.
Finally, we find it natural to impose global thermodynamic stability only where
local one exists for the given black hole solution.
}
\end{abstract}

\thispagestyle{empty}
\pagebreak
\tableofcontents

\section{Introduction}

In the past few decades investigating lower dimensional gravity theories has
become very attractive area of research. This is mainly due to the remarkable
gauge/gravity correspondence\cite{Maldacena:1997re}
which states a duality between particular gravitational and quantum systems.
A modern review of the correspondence can be found in \cite{ammon_erdmenger_2015}.
Within this framework, a number of $D = 3$ black hole solutions are shown to be
dual to two-dimensional quantum field theories at a finite temperature. The most
famous of these solutions is the Banados-Teitelboim-Zanelli (BTZ) black hole
\cite{Banados:1992wn} and its generalizations.

Until recently there were two ways of constructing three-dimensional models of
gravity. In the first approach one adds topological Chern-Simons terms to the
standard Einstein-Hilbert action \cite{deser:1982a, deser:1982b,
Bergshoeff:2014pca}. In the second approach the Einstein-Hilbert action is
modified by higher-derivative correction terms \cite{Bergshoeff:2009hq,
  Bergshoeff:2009aq}. Lately a third approach, involving higher-curvature
  corrections\footnote{Higher-curvature corrections are known to occur often in
    models of quantum gravity. There they arise precisely as quantum corrections to
    the Einstein-Hilbert action.} to the Einstein-Hilbert action, found its way
    down to three-dimensional gravity. It was shown that the $D > 5$
    Einstein-Gauss-Bonnet theory\footnote{Gauss-Bonnet gravity is the simplest
      representative of the Lovelock class of theories of gravity in higher
      than four spacetime dimensions. Lovelock gravity \cite{Lovelock:1971yv}
    maintains the property of having second-order field equations for all
  backgrounds.} possesses a non-trivial limit to four \cite{Glavan:2019inb,
Hennigar:2020lsl, Lu:2020iav} and lower spacetime dimensions \cite{Lu:2020iav, Hennigar:2020fkv,
Hennigar:2020drx}. The latter has been suggested to circumvent the Lovelock
theorem and allows the contribution of the higher-curvature Gauss-Bonnet term
to the local dynamics. While the proposed regularization procedure is not
consistent for general gravitational fields \cite{Gurses:2020ofy,Ai:2020peo,
Shu:2020cjw, Fernandes:2020nbq}, it leads to correct predictions in a number of
cases with high symmetries.

Although there are many gravitational solutions in three dimensions, to our
knowledge there exist very few novel $D=3$ Gauss-Bonnet black holes found in
\cite{Ma:2020ufk, Hennigar:2020fkv, Hennigar:2020drx}. The solutions given in
\cite{Ma:2020ufk, Hennigar:2020fkv} are Gauss-Bonnet generalization of the static BTZ
black hole with non-trivial scalar field profiles. The other solution is the
(2$+$1)-dimensional rotating Gauss-Bonnet black hole \cite{Hennigar:2020drx}. 

There are several issues, which motivate this study.
First of all, concerning the holographic conjecture, we expect the 3-dimensional rotating Gauss-Bonnet 
(RGB$_3$) black hole to be dual to a certain $2D$ CFT as in the case of the standard BTZ. In this context RGB$_3$ could also be related to a number of interesting phenomena such as SYK models \cite{PhysRevLettYe, Bulycheva:2019rtj},  holographic quantum matter \cite{Hartnoll:2016apf, zaanen:2015}, higher spin theory \cite{Klebanov:2002ja,Giombi:2009wh},  strongly-correlated lower-dimensional systems \cite{zaanen:2015}, etc. Secondly, the $D=4$ case has a clear physical content, however in a number of cases the system exhibits a reduction from $D\geq4$ to $D=3$ (see for instance \cite{Lu:2020iav, Hennigar:2020fkv}).
Since in $D=3$ the Gauss-Bonnet term $\mathcal{G}$ vanishes from the action, the connection to the Einstein-Gauss-Bonnet theory is seemingly lost\footnote{Note that in 3 dimensions Gauss-Bonnet terms vanishes, but the non-trivial profile of the scalar field keeps a contribution from the Gauss-Bonnet coupling $\alpha$.}. Therefore, the systematic analysis of the 3D Gauss-Bonnet black holes becomes an important issue, which is now qualitatively different from $D=4$ and higher-dimensional cases.

The goal of our
paper is to study the thermodynamic properties of the (2$+$1)-dimensional rotating Gauss-Bonnet black hole from local and global perspective. Our analysis can later be  transferred to the
dual quantum system. Similar analysis has already been conducted in
\cite{Dimov:2019fxp}, where one can constrain the dual left and right
central CFT charges using the bulk thermodynamics of the warped AdS$_3$ black
hole.

In standard black hole thermodynamics, one is interested only in the proper
state quantities -- those being the energy, the entropy, the charges and the
angular momenta of the black hole. That being said, black holes are thermal
systems -- this means that they might not necessarily be in thermodynamic
equilibrium with their environment. Thus, further considerations have to be
taken into account. Specifically, one discerns two types of equilibrium --
local and global.

If a given system is situated in a global thermodynamic equilibrium then, by definition,
it has the same temperature, the same pressure, the same chemical potentials
etc, everywhere in space. In this case, one can study the global thermodynamic
stability (GTDS) in a given ensemble by considering the properties of the
corresponding thermodynamic potentials.

The system is said to be in local thermodynamic equilibrium if one can divide
it into smaller constituents, which are individually in thermodynamic
equilibrium, at least approximately. These partial systems can also be
described by thermodynamic state quantities. However, it is of crucial
importance that the partial systems can be chosen large enough for
a statistical description to be reasonable. Nevertheless, in each partial
system the intensive thermodynamic state quantities assume definite constant
values and do not vary too strongly from one partial system to another, i.e.\
only small gradients are allowed. 

If the system is in local equilibrium, the local thermodynamic stability (LTDS)
does not imply a global one. On the other hand, it is natural to assume that
a system in global thermodynamic equilibrium is also locally stable.  Thus, it
is evident that GTDS always implies LTDS, but not vice versa. This notion will be
of topmost importance in our considerations.
Whilst analyzing the numerous specific heats definable in our extended
thermodynamic picture we will look for regions of intersection between LTDS and
GTDS\@. Only in such regions can one define true global thermodynamic equilibrium 
with respect to the corresponding specific heat.

The structure of the paper is as follows. In Section~\ref{secRGBBTZ} we present
the RGB$_3$ black hole solution and its thermodynamics.  In Section~\ref{sec3} we fix our equilibrium ensemble. Although the local
thermodynamic stability of black holes has been fully developed
\cite{Mansoori:2014oia}, to our knowledge the global thermodynamic stability
analysis has not been stated properly or in full for black holes. For that
reason in Section~\ref{sec4} we are going to  generalize the notion of global
thermodynamic stability from standard thermodynamics to describe the global
equilibrium of black holes. We will achieve this by applying a proper Legendre
transformation of the energy of the RGB$_3$ solution to find the
natural thermodynamic potential for the given ensemble of macro parameters. The global stability analysis, restricted to the 
	week positivity conjecture, is based on 
	the properties of the new thermodynamic potential. The advantage of this method is that it allows one to
chose different potentials, corresponding to different constraints to which the
system may be subjected. 
In Section~\ref{secLTDS} we investigate the local thermodynamic stability of
the RGB$_3$ black hole by analyzing the proper specific heats. As stated
previously, we find it natural
to impose global thermodynamic stability only where local one exists. Finally, in
Section~\ref{secConcl} we give our concluding remarks.

\section{The (2$+$1)-dimensional rotating Gauss–Bonnet black hole}\label{secRGBBTZ}

The generic Einstein-Gauss-Bonnet action in $D=d+1$ space-time dimensions is given by \cite{Lu:2020iav}:
\begin{equation}\label{key}
I=\frac{1}{16 \pi} \int d^D x \sqrt{|g|} \left[R-2 \Lambda+\alpha \left(\phi
\mathcal{G}+4 G^{\mu\nu}\partial_\mu\phi\partial_\nu\phi-4(\partial \phi)^2\Box
\phi+2 (\partial\phi)^4\right)\right],
\end{equation}
where\footnote{We also have the notations $\sqrt{|g|}\,\Box
\phi=\partial_\mu(\sqrt{|g|}g^{\mu\nu}\partial_\nu\phi)$,
$(\partial\phi)^2=g^{\mu\nu}\partial_\mu\phi\partial_\nu\phi$ and
$(\partial\phi)^4=(g^{\mu\nu}\partial_\mu\phi\partial_\nu\phi)^2$.}
$\phi(t,\vec x)$ is a scalar field,  $\Lambda=-{d(d-1)/(2\ell^2)}$ is the
cosmological parameter, $\alpha$ is the Gauss-Bonnet coupling and
$\mathcal{G}=R_{\mu\nu\rho\sigma} R^{\mu\nu\rho\sigma}-4 R_{\mu\nu}
R^{\mu\nu}+R^2$
is the Gauss-Bonnet term, which identically vanishes for $D<4$. Restricted to $D=2+1$ dimensions the above action becomes\footnote{Note that in 3 dimensions Gauss-Bonnet terms vanishes, but the non-trivial profile of the scalar field keeps a contribution from the Gauss-Bonnet coupling $\alpha$.} \cite{Lu:2020iav, Hennigar:2020fkv, Hennigar:2020drx}: 
\begin{equation}\label{key}
	I=\frac{1}{16 \pi} \int d^3 x \sqrt{|g|} \left[R-2 \Lambda+2\alpha\left(2 G^{\mu\nu}\partial_\mu\phi\partial_\nu\phi-2(\partial \phi)^2\Box
	\phi+ (\partial\phi)^4\right)\right].
\end{equation}

There exist a static BTZ black hole solution for this action given by the metric \cite{Hennigar:2020fkv, Ma:2020ufk}:
\begin{equation}\label{eqStaticGBBTZBH}
ds^2=-f dt^2+\frac{dr^2}{f}+r^2d\varphi^2,\quad \phi=\log{\frac{r}{l}}.
\end{equation}
Here $l$ is an integration constant and 
\begin{equation}\label{key}
f^{\pm}=-\frac{r^2}{2\alpha}\left(1\pm\sqrt{1+\frac{4\alpha}{r^2}f_E}\right)
\end{equation}
is the Gauss–Bonnet generalization of the (static) Einstein theory BTZ metric  \cite{Banados:1992wn}:
\begin{equation}\label{key}
f_E=\frac{r^2}{\ell^2}-m.
\end{equation}
In this case, only $f^{-}$ gives a black hole solution, which at  $\alpha\to0$
reduces to the standard Einstein BTZ with $R=-6/\ell^2$. Here, one also has the
cosmological length scale $\ell>0$ and two arbitrary integration constants $l$,
$m$. 

The $D=2+1$ rotating Gauss-Bonnet (RGB$_3$) black hole solution can be
obtained from (\ref{eqStaticGBBTZBH}) after performing the following boost
transformation on the coordinates \cite{Hennigar:2020drx}:
\begin{equation}\label{key}
  t\to \Xi t-a \varphi,\quad \varphi\to \frac{a t}{\ell^2}-\Xi\varphi,\quad
  \Xi_{}^2=1+\frac{a^2}{L^2_{}},\quad L=\sqrt{\frac{2
  \alpha}{\sqrt{1+\frac{4\alpha}{\ell^2}}-1}},\quad a\geq 0.
\end{equation}
Hence, the metric of the RGB$_3$ black hole yields
\begin{equation}\label{eqRGB3}
	ds^2=-f^{-} (\Xi dt-a d\varphi)^2+\frac{r^2}{L^4} (a dt-\Xi L^2 d\varphi)^2+\frac{dr^2}{f^{-}} ,\quad \phi=\log{\frac{r}{l}}.
\end{equation}

As in the static case, one has a black hole solution for $f^{-}$, which has
a proper limit at $\alpha\to0$. The inner Cauchy horizon is located at $r=0$
and the outer event horizon resides at $r_{h}=\ell\sqrt m$, where $m>0$.
Furthermore, the scalar curvature is
\begin{equation}\label{key}
R = \frac{{3{r^2}\left( {4\alpha  + {\ell ^2}} \right)\left( {r\ell \sqrt
X  - \left( {X - 2\alpha m{\ell ^2}} \right)} \right) - 12\alpha mr{\ell
^3}\sqrt X  - 16{\alpha ^2}{m^2}{\ell ^4}}}{{\alpha r\ell {X^{3/2}}}},
\end{equation}
where $X = {r^2}\left( {4\alpha  + {\ell ^2}} \right) - 4\alpha m{\ell ^2}$.
The curvature $R$ indicates physical singularities  at $r\to0$ and
\begin{equation}\label{key}
r_{cs} = \frac{{2\sqrt {m\alpha } \ell }}{{\sqrt {4\alpha  + {\ell ^2}} }}.
\end{equation}
When $\alpha>0$ the curvature singularity  is $r_{cs}>0$ and the metric
function cannot be extended all the way to $r=0$. When
$-\frac{\ell^2}{4}<\alpha<0$ the curvature singularity $r_{cs}$ is not real,
thus the metric function can be extended down to $r=0$. Finally, when
$\alpha<-\ell^2/4$,
the singularity at $r_{cs}$ reappears positive and real. This case, however,
corresponds to a naked singularity $r_{cs}>r_{h}$ and it will not be
considered.

\section{Extended thermodynamics and equilibrium space}\label{sec3}

\subsection{Extended thermodynamics}
The relevant thermodynamic state quantities of the RGB$_3$ black hole have 
already been obtained in \cite{Hennigar:2020drx}.  They can be written in the following way
\begin{equation}\label{eqOmega}
  T=  \frac{{\sqrt m }}{{2\pi \ell \sqrt {\frac{{{a^2}Y}}{{2\alpha }} + 1}
  }},\quad S = \frac{{\sqrt m \pi \ell }}{2}\sqrt {\frac{{{a^2}Y}}{{2\alpha }}
+ 1} ,
\end{equation}
\begin{equation}\label{eq:MSJ}
   \Omega  = \frac{{aY}}{{2\alpha \sqrt {\frac{{{a^2}Y}}{{2\alpha }} + 1} }}
   , \quad  J = \frac{{am}}{4}\sqrt {\frac{{{a^2}Y}}{{2\alpha }} + 1}, 
\end{equation}
\begin{equation}\label{key}
V = \pi m\left( {\frac{{{a^2}}}{{1 + Y}} + {\ell ^2}} \right),\quad
P=\frac{1}{8 \pi  \ell ^2},
\end{equation}
\begin{equation}\label{key}
M = \frac{m}{8}\left( {\frac{{{a^2}Y}}{\alpha } + 1} \right),\quad  {\Psi
} = \frac{{{a^2}m}}{{16{\alpha ^2}(1 + Y)}}\left( {Y - \frac{{2\alpha }}{{{\ell
^2}}}} \right),	
\end{equation}
where for convenience we have defined the parameter
\begin{equation}\label{key}
Y = \sqrt {\frac{{4\alpha }}{{{\ell ^2}}} + 1}  - 1.
\end{equation}

Furthermore, the parameter $P=-\Lambda/(8\pi)$ is proportional to the
cosmological constant and is interpreted as pressure \cite{Cvetic:2010jb}. Its conjugate thermodynamic
variable  $V$ is the thermodynamic volume of the black hole. The state quantity
$\Psi$ is the chemical potential for the Gauss-Bonnet parameter $\alpha$.
The first law of thermodynamics yields\cite{Hennigar:2020drx}:
\begin{equation}\label{eqFirstLawM}
		\delta M=T\delta S+ \Omega \delta J+V\delta P+\Psi\delta \alpha.
\end{equation}
Additionally, one has the Smarr relation
\begin{equation}\label{key}
 0=TS-2 PV+\Omega J+2 \Psi \alpha.
\end{equation}

In equilibrium, the standard relations between the intensive and the extensive
parameters hold:
\begin{equation}\label{key}
T=\frac{\partial M}{\partial S}\Big |_{J, P,\alpha},\quad \Omega=\frac{\partial
M}{\partial J}\Big |_{S, P,\alpha}, \quad V=\frac{\partial M}{\partial P}\Big
|_{S ,J,\alpha},\quad \Psi=\frac{\partial M}{\partial \alpha}\Big |_{S, J, P}.
\end{equation}
\\
In the limit $a\to0$ we recover the thermodynamics of the static Gauss-Bonnet
solution, which is identical to the Einstein BTZ black hole as
pointed out by \cite{Hennigar:2020drx}:
\begin{align}\label{eqTDNonRot}
M=\frac{m}{8},\quad T=\frac{\sqrt{m}}{2 \pi  \ell },\quad S=\frac{\pi\ell\sqrt m}{2},\quad P=\frac{1}{8 \pi \ell^2},\quad V=m \pi\ell^2,\quad \Psi=0,
\end{align}
with the first law and the Smarr relation reducing respectively to
\begin{equation}\label{key}
  \delta M=T\delta S+ V\delta P+\Psi\delta \alpha\quad\text{and}\quad 0=TS-2 PV+2 \Psi \alpha,
\end{equation}

It is natural to assume that certain physical parameters are always positive,
namely the mass, entropy, temperature and volume. It can be shown that
$M,S,T,V>0$ simultaneously lead to  
\begin{equation}\label{eqSectors}
\alpha>0 \quad \text{or}\quad -\ell^2/4<\alpha<0,
\end{equation}
thus confirming the two sectors of the solution. One can also consider
$\alpha\to 0^\pm$ and $\alpha\to -\ell^2/4$ as special cases, when it is
possible. Note that the naked singularity case $\alpha<-\ell^2/4$ has been
discarded. 

\subsection{The space of equilibrium states}
In order to mitigate some computational complexity in the study of the 
thermodynamics of the system we express ($M,S,J,V,\Psi$) in terms of the parameters
($T,\Omega,P,\alpha$). The later set will span our equilibrium space\footnote{It
becomes an equilibrium manifold after defining a proper Riemannian metric on
it, which is a case of study by the framework of thermodynamic information
geometry.}. To do so, one can solve $T$ and $\Omega$ (\ref{eqOmega},\ref{eq:MSJ}) for $m$ and $a$:
\begin{align}
m = \frac{{2{\pi ^2}{T^2}{\ell ^2}\left( {{a^2}Y + 2\alpha } \right)}}{\alpha
},\quad {a_ \pm } =  \pm \frac{{2\alpha \Omega }}{{\sqrt {Y\left( {Y - 2\alpha
{\Omega ^2}} \right)} }},
\end{align}
where we take $a_+>0$ for $\alpha>0$ and $a_->0$ for $-\ell^2/4<\alpha<0$. This
follows from the sign of $Y$, i.e.
\begin{equation}\label{key}
  Y = \sqrt {\frac{{4\alpha }}{{{\ell ^2}}} + 1}  - 1 = \sqrt {32\pi \alpha
    P + 1}  - 1\quad  \Rightarrow \quad \left\{ \begin{array}{l}
		Y > 0,\quad \alpha > 0,\\
		-1< Y < 0,\quad  - \frac{{{\ell ^2}}}{4} < \alpha < 0.
	\end{array} \right.
\end{equation}
One has to be careful with the condition $-1<Y<0$ , because $Y$ depends on
$\alpha$. The right and left bounds are $Y_1(\alpha, P)$ and $Y_2(\alpha,
P)$, which satisfy $Y_1(\alpha\to{-\ell^2/4}) \to-1$ and $Y_2(\alpha\to 0^-)
\to 0$. Therefore this condition actually looks like
\begin{equation}\label{key}
	-1< Y_1\leq Y\leq Y_2< 0.
\end{equation}

In both sectors for $\alpha$ the thermodynamics in ($T,\Omega,P,\alpha$)
equilibrium space takes the form:
\begin{align}&M = \frac{2 \pi ^2 \alpha  T^2 \left(2 \alpha
  \Omega ^2+Y\right)}{(Y+2) \left(Y-2 \alpha  \Omega ^2\right)^2},\quad
  S = \frac{4 \pi ^2 \alpha  T}{(Y+2) \left(Y-2 \alpha  \Omega ^2\right)},\quad
  J = \frac{8 \pi ^2 \alpha ^2 T^2 \Omega }{(Y+2) \left(Y-2 \alpha  \Omega
  ^2\right)^2},\nonumber\\
  &V = \frac{64 \pi ^3 \alpha ^2 T^2 \left(Y+1-\alpha  \Omega
  	^2\right)}{(Y+1) (Y+2)^2 \left(Y-2 \alpha  \Omega ^2\right)^2},\quad \Psi
  = -\frac{2 \pi ^2 \alpha  T^2 Y \Omega ^2}{\left(Y^2+3 Y+2\right) \left(Y-2
  \alpha  \Omega ^2\right)^2}. \label{eqTDparams} 
\end{align}

Assuming $T,P,\Omega>0$, all parameters in \eqref{eqTDparams} have a common
divergence in ($T,\Omega,P,\alpha$) space given by the following temperature
independent spinodal curve $Y-2 \alpha  \Omega ^2=0$.  It will be convenient to work with
$\omega = \Omega^2 >0$ instead of $\Omega$ throughout the paper, thus the spinodal curve can be written by
\begin{equation}\label{eqSpinodal}
s=\sqrt{32 \pi P \alpha +1}-1 -2 \alpha\omega=0,
\end{equation}
Solving $s=0$ with respect to $\omega$, one finds the following critical value:
\begin{equation}\label{eqCriticals}
	\omega_c=\frac{\sqrt{32 \pi  \alpha  P+1}-1}{2 \alpha } = \frac{Y}{2\alpha}.
\end{equation}
One notes that $\omega_c >0$ holds in both sectors for $\alpha$. Furthermore, since
the entropy from (\ref{eqTDparams}) has to be positive, $S>0$, it is evident
that
the following restriction on $\omega$ must hold
\begin{equation}
	\frac{\alpha}{Y-2\alpha\omega}>0,
\end{equation}
which reduces to
\begin{equation}
	\omega<\omega_c.
\end{equation}
Hence, all physically meaningful states occur for values of $\omega$ less than
the critical value $\omega_c$. No physical states exist for $\omega>\omega_c$. The
analysis of the thermodynamic properties of the system, close to the spinodal
curve $s=0$  $(\omega=\omega_c)$ requires the methods of non-equilibrium
thermodynamics. We leave this path of inquiry for future work.

The critical squared angular velocity $\omega_c$ is a decreasing function of
$\alpha$. For positive values of $\alpha$ this parameter is bounded from above and below,
i.e.
\begin{equation}\label{key}
	0<\omega_c<8\pi P,\quad \quad \alpha> 0,
\end{equation}
which follows from the limits $\lim\limits_{\alpha\to 0^+}\omega_c = 8\pi P$
and $\lim\limits_{\alpha\to \infty}\omega_c = 0$. The negative values of
$\alpha$ are bounded from below $\alpha_p< \alpha$, where
$\alpha_p=-\ell^2/4=-1/(32 \pi  P)$ is the physical lower bound. In this case,
the parameter $\omega_c$ is bounded from both sides:
\begin{equation}\label{key}
	8\pi P < \omega_c < 16\pi P,\quad \alpha_p<\alpha<0,
\end{equation}
where $\lim\limits_{\alpha\to \alpha_p}\omega_c = 16\pi P$.

In what follows we will investigate the local and global thermodynamic
properties of the ($2+1$)-dimensional rotating Gauss-Bonnet solution (\ref{eqRGB3}) in both sectors for
$\alpha$ in ($T,\Omega,P,\alpha$) equilibrium space.

\section{Global thermodynamic stability}\label{sec4}

In this section we present our approach to the  global thermodynamic stability of the RGB$_3$ black hole solution. Our investigation is conducted entirely within the weak positivity conjecture of the Hessian of the corresponding thermodynamic potential.

\subsection{Weak and strong global conditions for thermodynamic stability}

One can study the global thermodynamic stability in a given ensemble by
considering the properties of the corresponding thermodynamic potentials. In
thermodynamics it is conventional to begin with the energy potential (for other potentials see Appendices \ref{appAB} and \ref{appA}). In the
extended black hole thermodynamics  the mass does not coincide with the energy
of the black hole, but it is interpreted as the enthalpy of spacetime:
\begin{equation}\label{key}
	M=E+P V=H.
\end{equation}
If we take the differential from both sides of this equation and solve for
$dE$, after comparing with the first law (\ref{eqFirstLawM}), we find
\begin{equation}\label{eqFirstLawE}
	\delta E=T\delta S+ \Omega \delta J-P\delta V+\Psi\delta \alpha.
\end{equation}
The first law in this form determines the natural parameters for the energy
potential, i.e. the energy is a function of $E=E(S, J, V, \alpha)$, where the natural parameters $(S, J, V,\alpha)$ are called energy extensive\footnote{Their conjugate $(T, \Omega, P,\Psi)$ are called energy intensive parameters.}.
On the other hand, the energy of the system is globally convex in its natural parameters, hence
\begin{equation}\label{eqGlobalCondE}
	\frac{\partial^2 E}{\partial S^2}\Big |_{J, V, \alpha}\geq 0,
	\quad \frac{\partial^2 E}{\partial J^2}\Big |_{S, V, \alpha}\geq 0,
	\quad \frac{\partial^2 E}{\partial V^2}\Big |_{S, J, \alpha}\geq 0,
	\quad \frac{\partial^2 E}{\partial \alpha^2}\Big |_{S,J,V }\geq 0.
\end{equation}
These conditions can be regarded as the weak global conditions for thermodynamic stability (see for example \cite{riseborough2020statistical}). However, in equilibrium the energy of the system should be
in its minimum, therefore the Hessian matrix of the energy should be positive semi-definite. The latter is determined by the Sylvester criterion, which states that all principle minors of the Hessian should be non-negative. Hence, the Sylvester criterion defines the necessary and sufficient conditions for global thermodynamic stability during various thermodynamic processes. 

In order to show how this works for the energy potential in its natural parameters let us consider a process, where only the entropy is allowed to fluctuate and keep all other state quantities fixed. In this case the first law (\ref{eqFirstLawE}) reduces to
\begin{equation}\label{key}
		\delta E=T\delta S.
\end{equation}
Therefore, for achieving a minimum of the energy (GTDS), one requires only the condition 
\begin{equation}\label{key}
	\Delta_{J,V,\alpha}=\frac{\partial^2 E}{\partial S^2}\Big|_{J,V,\alpha}\geq 0.
\end{equation}
Similarly, if one allows only for the angular momentum to fluctuate, the condition for GTDS becomes
\begin{equation}\label{key}
\Delta_{S,V,\alpha}=\frac{\partial^2 E}{\partial J^2}\Big|_{S,V,\alpha}\geq 0.
\end{equation}
The same reasoning holds for fluctuations along $V$ or $\alpha$, leading to the third and fourth condition in (\ref{eqGlobalCondE}). As a result one can interpret the conditions for weak global thermodynamic stability (\ref{eqGlobalCondE}) as processes, where the state quantities (the natural parameters of the energy) fluctuate individually and independently of each other.

Now, let us consider processes with two of the state quantities  fluctuating simultaneously. In this case, the Sylvester criterion implies that the weak conditions from (\ref{eqGlobalCondE}) should still hold together with
\begin{align}\label{eqStrongerTD}
\nonumber	&\Delta_{S,J}=\left|
	\begin{array}{cc}
		\frac{\partial^2 E}{\partial V^2} & \frac{\partial^2 E}{\partial V \partial \alpha} \\[5pt]
		\frac{\partial^2 E}{\partial V\partial \alpha} & \frac{\partial^2 E}{\partial \alpha^2}\\
	\end{array}
	\right|_{S,J}\geq 0, 
	\quad \Delta_{S,V}=\left|
	\begin{array}{cc}
		\frac{\partial^2 E}{\partial J^2} & \frac{\partial^2 E}{\partial J \partial \alpha} \\[5pt]
		\frac{\partial^2 E}{\partial J\partial \alpha} & \frac{\partial^2 E}{\partial \alpha^2}\\
	\end{array}
	\right|_{S,V}\geq 0, 
\nonumber	\\
		&\Delta_{S,\alpha}=\left|
	\begin{array}{cc}
		\frac{\partial^2 E}{\partial J^2} & \frac{\partial^2 E}{\partial J \partial V} \\[5pt]
		\frac{\partial^2 E}{\partial J\partial V} & \frac{\partial^2 E}{\partial V^2}\\
	\end{array}
	\right|_{S,\alpha}\geq 0, 
	\quad \Delta_{J,V}=\left|
	\begin{array}{cc}
		\frac{\partial^2 E}{\partial S^2} & \frac{\partial^2 E}{\partial S \partial \alpha} \\[5pt]
		\frac{\partial^2 E}{\partial S\partial \alpha} & \frac{\partial^2 E}{\partial S^2}\\
	\end{array}
	\right|_{J,V}\geq 0, 
\nonumber	\\
		&\Delta_{J,\alpha}=\left|
	\begin{array}{cc}
		\frac{\partial^2 E}{\partial S^2} & \frac{\partial^2 E}{\partial S \partial V} \\[5pt]
		\frac{\partial^2 E}{\partial S\partial V} & \frac{\partial^2 E}{\partial V^2}\\
	\end{array}
	\right|_{J,\alpha}\geq 0, 
	\quad \Delta_{V,\alpha}=\left|
	\begin{array}{cc}
		\frac{\partial^2 E}{\partial S^2} & \frac{\partial^2 E}{\partial S \partial J} \\[5pt]
		\frac{\partial^2 E}{\partial S\partial J} & \frac{\partial^2 E}{\partial J^2}\\
	\end{array}
	\right|_{V,\alpha}\geq 0.
\end{align}
Here, the $\Delta_{i,j}$ are the principal minors of the Hessian (\ref{eqStrongerTDLevel4}) with two rows and columns removed, corresponding to the fixed parameters.

If one allows for three fluctuating state quantities the conditions (\ref{eqGlobalCondE}) and (\ref{eqStrongerTD}) should hold together with
\begin{align}\label{eqStrongerTDLevel3}
\nonumber &\Delta_S=\left|
\begin{array}{ccc}
	\frac{\partial^2 E}{\partial J^2} & \frac{\partial^2 E}{\partial J \partial V} &\frac{\partial^2 E}{\partial J \partial \alpha}\\[5pt]
	\frac{\partial^2 E}{\partial J\partial V} & \frac{\partial^2 E}{\partial V^2} & \frac{\partial^2 E}{\partial V \partial \alpha}\\[5pt]
	\frac{\partial^2 E}{\partial J\partial \alpha} & \frac{\partial^2 E}{\partial V\partial \alpha} & \frac{\partial^2 E}{ \partial \alpha^2}
\end{array}
\right|_S \geq 0,
\quad \Delta_J=\left|
\begin{array}{ccc}
	\frac{\partial^2 E}{\partial S^2} & \frac{\partial^2 E}{\partial S \partial V} &\frac{\partial^2 E}{\partial S \partial \alpha}\\[5pt]
	\frac{\partial^2 E}{\partial S\partial V} & \frac{\partial^2 E}{\partial V^2} & \frac{\partial^2 E}{\partial V \partial \alpha}\\[5pt]
	\frac{\partial^2 E}{\partial S\partial \alpha} & \frac{\partial^2 E}{\partial V\partial \alpha} & \frac{\partial^2 E}{ \partial \alpha^2}
\end{array}
\right|_J \geq 0,
\\
&\Delta_V=\left|
\begin{array}{ccc}
	\frac{\partial^2 E}{\partial S^2} & \frac{\partial^2 E}{\partial S \partial J} &\frac{\partial^2 E}{\partial S \partial \alpha}\\[5pt]
	\frac{\partial^2 E}{\partial S\partial J} & \frac{\partial^2 E}{\partial J^2} & \frac{\partial^2 E}{\partial J \partial \alpha}\\[5pt]
	\frac{\partial^2 E}{\partial S\partial \alpha} & \frac{\partial^2 E}{\partial J\partial \alpha} & \frac{\partial^2 E}{ \partial \alpha^2}
\end{array}
\right|_V \geq 0,
\quad \Delta_\alpha=\left|
\begin{array}{ccc}
	\frac{\partial^2 E}{\partial S^2} & \frac{\partial^2 E}{\partial S \partial J} &\frac{\partial^2 E}{\partial S \partial V}\\[5pt]
	\frac{\partial^2 E}{\partial S\partial J} & \frac{\partial^2 E}{\partial J^2} & \frac{\partial^2 E}{\partial J \partial V}\\[5pt]
	\frac{\partial^2 E}{\partial S\partial V} & \frac{\partial^2 E}{\partial J\partial V} & \frac{\partial^2 E}{ \partial V^2}
\end{array}
\right|_\alpha \geq 0,
\end{align}
where the $\Delta_{i}$ are the principal minors of the Hessian (\ref{eqStrongerTDLevel4}) with one row and column removed.

Finally, if all state quantities are allowed to fluctuate simultaneously, one has an additional condition, namely the determinant of the Hessian to be non-negative:
\begin{equation}\label{eqStrongerTDLevel4}
	\Delta=\left|
	\begin{array}{cccc}
		\frac{\partial^2 E}{\partial S^2} & \frac{\partial^2 E}{\partial S \partial J} &\frac{\partial^2 E}{\partial S \partial V}&\frac{\partial^2 E}{\partial S \partial \alpha}\\[5pt]
		\frac{\partial^2 E}{\partial S\partial J} & \frac{\partial^2 E}{\partial J^2} & \frac{\partial^2 E}{\partial J \partial V}& \frac{\partial^2 E}{\partial J \partial \alpha}\\[5pt]
		\frac{\partial^2 E}{\partial S\partial V} & \frac{\partial^2 E}{\partial J\partial V} & \frac{\partial^2 E}{ \partial V^2}&\frac{\partial^2 E}{\partial V\partial \alpha}\\[5pt]
		\frac{\partial^2 E}{\partial S\partial \alpha} & \frac{\partial^2 E}{\partial J\partial \alpha} & \frac{\partial^2 E}{ \partial V\partial\alpha}&\frac{\partial^2 E}{\partial \alpha^2}
	\end{array}
	\right|\geq 0.
\end{equation}

The conditions (\ref{eqGlobalCondE}), (\ref{eqStrongerTD}), (\ref{eqStrongerTDLevel3}) and (\ref{eqStrongerTDLevel4}) determine the strong global thermodynamic stability of the RGB$_3$ black hole with respect to the energy in its natural parameters. However, we have a solution for the energy potential in ($T,\Omega,P,\alpha$) space, 
\begin{equation}\label{EnergyPot}
	E(T,\Omega,P,\alpha)=M-P V=\frac{\pi  \alpha  T^2 Y (3 Y+2) \Omega ^2}{16 P (Y+1) \left(Y-2 \alpha  \Omega ^2\right)^2}.
\end{equation}
This shows that the energy is not the appropriate thermodynamic potential in ($T,\Omega,P,\alpha$) space, because it is not a function of its natural parameters and therefore the above conditions for GTDS are not applicable. In order to find the appropriate thermodynamic potential,
whose natural state parameters are ($T,\Omega,P,\alpha$),  one has to perform Legendre transformation of the energy from  $(S, J, V, \alpha)$ space
to some new potential $\Phi$ in $(T,\Omega,P,\alpha)$ space, i.e.
\begin{equation}\label{key}
	\Phi(T,\Omega,P,\alpha)=\mathcal{L}_{S, J, V} E=E-T S-\Omega J+P V= -\frac{\pi  T^2 Y}{16 P Y-32 \alpha  P \Omega ^2},
\end{equation}
where the subscripts $S,J,V$ of $\mathcal{L}_{S,J,V}$ indicate the parameters, which the Legendre transformation is applied on.
The first law of thermodynamics now reads
\begin{equation}\label{key}
	\delta\Phi=-S \delta T-J \delta \Omega+V \delta P+\Psi \delta \alpha.
\end{equation}
This confirms that the natural state parameters for $\Phi$ are exactly
$(T,\Omega,P,\alpha)$. As a result of the Legendre transformation the new potential $\Phi$ is now concave along $(T,\Omega,P)$ and convex along $\alpha$:
\begin{equation}\label{eqGlobalPhi}
	\frac{\partial^2 \Phi}{\partial T^2}\Big |_{ \Omega,P, \alpha}\leq 0,
	\quad \frac{\partial^2 \Phi}{\partial \Omega^2}\Big |_{T, P, \alpha}\leq 0,
	\quad \frac{\partial^2 \Phi}{\partial P^2}\Big |_{T, \Omega, \alpha}\leq 0,
	\quad \frac{\partial^2 \Phi}{\partial \alpha^2}\Big |_{T,\Omega,P }\geq 0.
\end{equation}
The flip of the signs of the first three conditions in (\ref{eqGlobalPhi}), as
compared to the conditions of the energy
(\ref{eqGlobalCondE}), is due to the fact that the product of the corresponding
conjugate thermodynamic variables in the Legendre transformation is always
taken as minus. One notes that the potential $\Phi$ is not strictly concave or convex, thus the Sylvester criterion is not directly applicable here. As a consequence, in the new ensemble $(T,\Omega,P,\alpha)$, one can only impose the weak global conditions (\ref{eqGlobalPhi}) for $\Phi$. In Appendix \ref{appAB} we show how to obtain the weak GTDS conditions for other thermodynamic potentials as well. In general, the procedure for obtaining the strong global conditions (a proper Sylvester criterion) for the new Legendre transformed potentials is subtle and will be presented in a separate survey. 

In what follows we are going to investigate the weak global thermodynamic stability for the potential $\Phi$ of the RGB$_3$ black hole solutions in $(T,\Omega,P,\alpha)$ space.

\subsection{Positive Gauss–Bonnet parameter, $\alpha>0$}\label{subsecPosA}

We can explicitly express conditions (\ref{eqGlobalPhi}) with respect to the relevant
parameters. For example, we choose to solve simultaneously (\ref{eqGlobalPhi})
with respect to $\omega$. In this case, the GTDS conditions reduce
to\footnote{The index $g$ in $\omega_g$ stands for ``global''.}
\begin{equation}\label{eqGlobalExplicitA}
0< \omega \leq \omega_g,
\end{equation}
where
\begin{equation}\label{key}
\omega_g=\frac{Y+2}{3Y+4}\omega_c=\frac{2 \pi  P \left(-1+3 \sqrt{32 \pi  \alpha  P+1}\right)}{36 \pi  \alpha  P+1}.
\end{equation}
Noticing that $(Y+2)/(3Y+4) <1$, it follows that $\omega\leq \omega_g<
\omega_c$ for all values of $P$ and $\alpha$ in this sector.
Therefore, the global spinodal $\omega_c$ is a boundary of the region of global
thermodynamic stability. The hierarchy between different $\omega$ in this
sector is
\begin{equation}\label{}
0<\omega \leq \omega_g <\omega_c<8\pi P.
\end{equation}
Furthermore, the upper global bound $\omega_g$ never intersects with
$\omega_c$, unless  $P\to 0$, where $\omega_g=\omega_c=0$. The GTDS in this
sector is depicted in Fig.~\ref{FigGTDSApos}. We also note that $\omega$ can become grater than $\omega_g$ for LTDS.

\subsection{Negative Gauss–Bonnet parameter, $\alpha_p< \alpha <0$}\label{secGTDSneg}

In this sector the Gauss-Bonnet parameter is negative and bounded from bellow
$\alpha_p<\alpha <0$. The condition for GTDS from Eq. \eqref{eqGlobalPhi} lead
to couple of distinct cases. We keep in mind that in all cases
$\omega_g<\omega_c$ holds.

\begin{figure}[h] 
	\begin{center}
		\begin{subfigure}{0.3\textwidth}
		\begin{tikzpicture}
			\draw[-{Latex},thick] (0,0) -- (3.7,0);
			\foreach \x in  {0,1.7,2.8} \draw[shift={(\x,0)},color=black, thick] (0pt,3pt) -- (0pt,-3pt);
			\foreach \x in {0} \draw[shift={(\x,0)},color=black] (0pt,0pt) -- (0pt,-3pt) node[below] {$\x$};
			\filldraw (1.7,-0.15) node[align=left,   below] {$\omega_g$};
			\filldraw (2.8,-0.15) node[align=left, below] {$\omega_c$};
			\filldraw (3.7,-0.15) node[align=left, below] {$\omega$};
			\draw[red,thick] (0,0).. controls +(up:0.43cm) and +(left:0cm)..
			node[above,sloped] {} (0.4,0.4);
			\draw[red,thick] (0.4,0.4)--(1.71,0.4);
			\draw[red,thick] (1.7,0.4)--(1.7,0);
			\end{tikzpicture}
			\caption{$\alpha>0$.}\label{FigGTDSApos}
		\end{subfigure}
		\begin{subfigure}{0.3\textwidth}
			\begin{tikzpicture}
				\draw[-{Latex},thick] (0,0) -- (3.7,0);
				\foreach \x in  {0,1.7,2.8} \draw[shift={(\x,0)},color=black, thick] (0pt,3pt) -- (0pt,-3pt);
				\foreach \x in {0} \draw[shift={(\x,0)},color=black] (0pt,0pt) -- (0pt,-3pt) node[below] {$\x$};
				\filldraw (1.7,-0.15) node[align=left,   below] {$\omega_g$};
				\filldraw (2.8,-0.15) node[align=left, below] {$\omega_c$};
				\filldraw (3.7,-0.15) node[align=left, below] {$\omega$};
				\draw[red,thick] (0,0).. controls +(up:0.43cm) and +(left:0cm)..
				node[above,sloped] {} (0.4,0.4);
				\draw[red,thick] (0.4,0.4)--(1.71,0.4);
				\draw[red,thick] (1.7,0.4)--(1.7,0);
			\end{tikzpicture}
			\caption{$-\frac{1}{36\pi P}\leq \alpha< 0$.}\label{FigGTDSAneg1}
		\end{subfigure}
		\begin{subfigure}{0.3\textwidth}
			\begin{tikzpicture}
				\draw[-{Latex},thick] (0,0) -- (4.5,0);
				\foreach \x in  {0,1.2,2,3,3.8} \draw[shift={(\x,0)},color=black, thick] (0pt,3pt) -- (0pt,-3pt);
				\foreach \x in {0} \draw[shift={(\x,0)},color=black] (0pt,0pt) -- (0pt,-3pt) node[below] {$\x$};
				\filldraw (1.2,-0.15) node[align=left,   below] {$\omega_+$};
				\filldraw (2,-0.15) node[align=left,   below] {$\omega_-$};
				\filldraw (3,-0.15) node[align=left,   below] {$\omega_g$};
				\filldraw (3.8,-0.15) node[align=left, below] {$\omega_c$};
				\filldraw (4.5,-0.15) node[align=left, below] {$\omega$};
				\draw[red,thick] (0,0).. controls +(up:0.45cm) and +(left:0cm)..
				node[above,sloped] {} (0.4,0.4);
				\draw[red,thick] (0.5,0.4)--(1,0.4);
				\draw[red,thick] (0.4,0.4)--(1.21,0.4);
				\draw[red,thick] (1.2,0.4)--(1.2,0);
				\draw[red,thick] (2,0)--(2,0.4);
				\draw[red,thick] (1.99,0.4)--(3.01,0.4);
				\draw[red,thick] (3,0.4)--(3,0);
			\end{tikzpicture}
			\caption{$\alpha_p< \alpha\leq -\frac{3}{100\pi P}$.}\label{FigGTDSAneg2}
		\end{subfigure}
	\end{center}
	\vspace{-10pt}
  \caption{Intervals of global thermodynamic stability: a) GTDS for $\alpha>0$ occurs in the interval
  $0<\omega\leq\omega_g$; b) GTDS for $-1/(36\pi P)\leq\alpha< 0$ occurs in the interval $0<
\omega\leq\omega_g$. c) GTDS for $\alpha_p<\alpha\leq-3/(100\pi P)$ occurs in
$0<\omega\leq\omega_{+}$ or $\omega_{-}\leq \omega\leq\omega_{g}$. The point
$\omega_{+}$ coincides with $\omega_{-}$ when $\alpha=-3/100\pi P$. In this
case the left and right GTDS merge together.}\label{FigGTDS}
\end{figure}
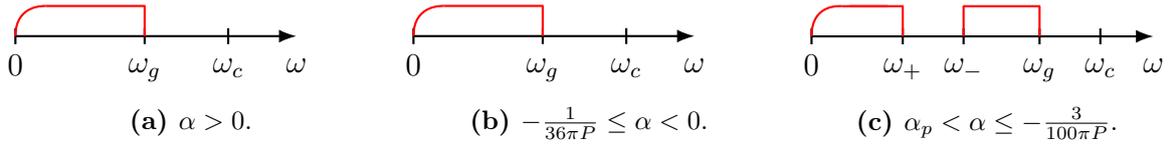

\begin{itemize}
	\item The simplest GTDS case corresponds to:
\begin{equation}\label{key}
	 0<\omega\leq \omega_g ,\qquad -\frac{1}{36\pi P}\leq \alpha < 0.
\end{equation}
This situation is shown on Fig.~\ref{FigGTDSAneg1}.
\item 
The second GTDS case is valid for $\alpha_p<\alpha<-\frac{1}{36\pi P}$. It
divides in two disjoint cases, where a more strict condition for $\alpha$
emerges:
\begin{equation}\label{eqGTDS2}
0<\omega\leq \omega_+, \quad \text{or} \quad \omega_-\leq\omega\leq\omega_g, \quad \text{where} \quad \alpha_p<\alpha\leq-\frac{3}{100\pi P},
\end{equation}
with $\omega_{\pm}$ given by
\begin{equation}\label{key1}
	\omega_\pm=\frac{9 Y(Y+2)+8 \pm (Y+2) \sqrt{-(3 Y+4) (5 Y+4)}}{4 \alpha  (3 Y+4)}.
\end{equation}
This case is pictured on Fig.~\ref{FigGTDSAneg2}.
\end{itemize}

Let us shortly discuss the results for the global thermodynamic stability of
the RGB$_3$ black hole. In the $\alpha>0$ sector there is only one condition
for GTDS given in Eq. (\ref{eqGlobalExplicitA}). The situation in the
$\alpha_p<\alpha<0$ sector is more complicated. Here one has two distinct
cases for GTDS of the black hole. However, when considering the global
thermodynamic stability together with the local one, we have to take into
account only the intersections between both types of stabilities in order to
have true GTDS\@. The reason for this follows form the fact that while LTDS
does not require GTDS,
the GTDS always implies LTDS\@. This analysis will be conducted in Section~\ref{secLTDS}.

\section{Local thermodynamic stability}\label{secLTDS}

The local thermodynamic stability (LTDS) of the RGB$_3$ black
hole can be determined by investigating the properties of the corresponding specific heats. 
The direct way could be to just take the derivative of the entropy with respect
to the temperature, which will result in the following heat capacity
\begin{equation}\label{eqDirectHC}
	C= T\left(\frac{\partial S}{\partial T}\right) = \frac{4 \pi ^2 \alpha  T}{(Y+2) \left(Y-2 \alpha  \Omega ^2\right)},
\end{equation}
which coincides with the entropy in $(T,\Omega,P,\alpha)$ space.
The problem in multi-parameter thermodynamic systems, such as the RGB$_3$ black
hole, is that there are multiple specific heats to choose from, because one has to keep track of which state quantities are fixed when calculating
 the specific heats. In this case we can
refer to the Nambu bracket formalism developed by \cite{Mansoori:2014oia}. Lets start by an ensemble with parameters
$(A,B,C,D)$, hence, for specific heat with constant parameters
$(E,F,G)$, the following relation holds:
\begin{equation}
  C_{E,F,G} = T\left(\frac{\partial S}{\partial T}\right)_{E,F,G}
  = T\frac{\left\{S,E,F,G\right\}_{A,B,C,D}}{\left\{T,E,F,G\right\}_{A,B,C,D}}.
\end{equation}
In our case $(A,B,C,D)=(T,\Omega,P,\alpha)$ define the parameters of the
ensemble, and $(E,F,G)$ span all the other thermodynamic quantities
$(\Omega,J,V,P,\Psi,\alpha)$, with  $\Phi$ being the thermodynamic potential.
Therefore, the relevant heat capacities for the RGB$_3$ solutions are (see
Appendix~\ref{appB}):
\begingroup
\addtolength{\jot}{1em}
\begin{align}\label{eq:specificheats}
& C_{\Omega, P, \alpha}=\frac{4\pi^2 T\alpha}{(Y+2)(Y-2\alpha\omega)},\\
&C_{J, P,\alpha}=\frac{4 \pi^2 T \alpha}{(Y+2)(Y+6\alpha\omega)},\\
& C_{\Omega, P, \Psi}=\frac{4\pi^2 T\alpha}{Y(Y+2)-2\alpha(3Y+4)\omega},\\
&C_{J, P,\Psi}=\frac{4 \pi ^2 \alpha  T (3 Y+4)}{(Y+2) \left(2 \alpha  \omega (7 Y+10)+Y (3 Y+4)\right)},\\
& C_{\Omega, V, \alpha}=\frac{4 \pi ^2 T \alpha ^2  \omega}{\alpha \omega (3 Y+4)  \left(2+3 Y-2 \alpha  \omega\right)-4 (Y+1)^3},\\
& C_{\Omega, V ,\Psi}=\frac{4 \pi ^2 \alpha ^2 T \omega}{(Y+2) \left(Y-2 \alpha  \omega\right) \left(\alpha \omega+Y (Y+1) \left(2+Y-2 \alpha  \omega\right)\right)},\\
&C_{J, V,\alpha}=\frac{4 \pi ^2 \alpha ^2 T (3 Y+4) \omega}{(Y+2) \left(\alpha  \omega \left(-5 Y^2+2 \alpha  (7 Y+10) \omega-12 Y-8\right)-4 (Y+1)^3\right)},\\
&C_{J, V,\Psi}=\frac{4 \pi ^2 \alpha ^2 T (1-2 Y (Y+1)) \omega}{(Y+2)
\left(\alpha  \omega \left(Y \left(2 Y^2+6 Y+5\right)-2 \alpha  (2 Y-1) (2 Y+3)
\omega\right)+(Y+1) (3 Y+4) Y^2\right)}.
\end{align}
\endgroup

One notes that the heat capacity from Eq. (\ref{eqDirectHC}) now corresponds to
$C_{\Omega, P, \alpha}$. The latter has only one singular curve $Y=2 \alpha  \omega$,
matching the spinodal (\ref{eqSpinodal}) of the natural global thermodynamic
potential $\Phi$ in $(T,\Omega,P,\alpha)$ space.

In general, if a given specific heat is positive then the system is
thermodynamically stable from local standpoint with respect to this specific
heat. If the corresponding specific heat is negative -- the system is not in
a local equilibrium. Finally, if the specific heat changes sign or diverges it
indicates phase transitions of the system.

Let us now study
the local thermodynamic stability of the RGB$_3$ black hole.

\subsection{Positive Gauss–Bonnet parameter, $\alpha >0$}

\subsubsection{Specific heat $C_{\Omega, P,\alpha}$}

\begin{figure}[h]
\begin{center}
	\begin{tikzpicture}
		\draw[-{Latex},thick] (0,0) -- (5,0);
		\foreach \x in  {0,2.2,3.5} \draw[shift={(\x,0)},color=black, thick] (0pt,3pt) -- (0pt,-3pt);
		\foreach \x in {0} \draw[shift={(\x,0)},color=black] (0pt,0pt) -- (0pt,-3pt) node[below] {$\x$};
		\filldraw (4.9,-0.15) node[align=left,   below] {$\omega$};
		\filldraw (2.2,-0.15) node[align=left,   below] {$\omega_g$};
        \filldraw (3.5,-0.15) node[align=left, below] {$\omega_c$};
		\draw[red,thick] (0.015,0).. controls +(up:0.43cm) and +(left:0cm)..
		node[above,sloped] {} (0.41,0.4);
		\draw[red,thick] (0.41,0.4)--(2.21,0.4);
		\draw[red,thick] (2.2,0)--(2.2,0.4);
		\draw[blue,thick] (0.7,0.7)--(2.8,0.7);
		\draw[blue,thick] (-0.01,0).. controls +(up:0.75cm) and +(left:0cm)..
		node[above,sloped] {} (0.7,0.7);
		\draw[blue,thick] (3.5,0).. controls +(up:0.75cm) and +(left:0cm)..
		node[above,sloped] {} (2.8,0.7);
	\end{tikzpicture}
\end{center}
\caption{Intersection intervals of GTDS (red curve) and LTDS (blue curve) for fixed
$(\Omega, P,\alpha)$.The same situation occurs also for fixed $(J, P, \alpha)$,
$(J,P,\Psi)$ and $(\Omega, V,\Psi)$. }\label{figHeat1} 
\end{figure}
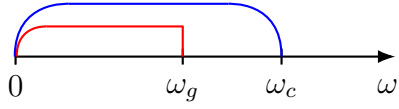

The first heat capacity we will consider in $(T,\Omega,P,\alpha)$ space is
$C_{\Omega, P,\alpha}$. It has one divergence at $Y=2\alpha \omega$, which
corresponds to the global spinodal curve $\omega_c$ for the potential $\Phi$.
The heat capacity $C_{\Omega, P,\alpha}$ is positive for  $0<\omega<\omega_c$.
The latter defines the region of local thermodynamic stability for the RGB$_3$ 
black hole with respect to fixed $(\Omega, P,\alpha)$. The global stability from Eq.
\eqref{eqGlobalExplicitA} falls within $0<\omega<\omega_g$, and it is fully covered by the LTDS,  due to the fact that
$\omega_g<\omega_c$. The situation is shown on Fig.~\ref{figHeat1}. 

Furthermore, several limiting cases occur for this heat capacity at $\alpha\to 0^{+}$, $\Omega\to 0$ and $P\to 0$:
\begin{equation}\label{key}
	C_{\Omega, P,\,\alpha\to 0^{+}}=\frac{\pi ^2 T}{8 \pi  P-\Omega ^2},
	\quad 	C_{\Omega\to 0,\, P,\,\alpha}=\frac{\pi  T}{8 P},
	\quad C_{\Omega,\, P\to 0,\,\alpha}=-\frac{\pi ^2 T}{\Omega ^2}.
\end{equation}
These limits indicate a transition to different black hole solutions. For example, the limit $\alpha\to 0^+$ leads to the rotating BTZ case, where one has $C_{\Omega, P,\alpha\to 0^{+}}>0$,
if only $\omega<8\pi P$, and  negative $C_{\Omega, P,\alpha\to 0^{+}}<0$ for
$\omega>8\pi P$. In the static case, $\Omega\to 0$, one finds that $C_{\Omega\to 0,\, P,\,\alpha}>0$ is
always positive. In the non-extended case, $P\to 0$, the heat capacity $C_{\Omega,\, P\to 0,\,\alpha}<0$ is
always negative.

\subsubsection{Specific heat $C_{J, P,\alpha}$}
The next specific heat $C_{J, P,\alpha}$ has no occurring divergences and is
always positive in this sector. Thus the RGB$_3$  black hole is locally  stable
at constant $(J, P,\alpha)$ in the physical region $0<\omega<\omega_c$ (see
Fig.~\ref{figHeat1}).
The three limiting cases here are:
\begin{equation}\label{key}
C_{J, P,\alpha\to 0^{+}}=\frac{\pi ^2 T}{8 \pi  P+3 \Omega ^2},\quad C_{J
, P,\alpha}\mathop{=}_{\Omega\to 0}\frac{\pi  T}{8 P}, \quad C_{J, P\to
0,\alpha}=\frac{\pi ^2 T}{3 \Omega ^2},
\end{equation}
all of which are positive.

\subsubsection{Specific heat $C_{\Omega, P, \Psi}$}

For constant $(\Omega, P, \Psi)$ the specific heat $C_{\Omega, P, \Psi}$ is
positive in the interval $0<\omega<\omega_g$, where $\omega_g$ is defined in Eq.
\eqref{eqGlobalExplicitA} and corresponds to a divergence in the heat capacity.
In this case the GTDS coincides with the LTDS with
$\omega_g$ not included in GTDS\@. Furthermore,
the singular curve $\omega_g$ and the global spinodal $\omega_{c}$ never
intersect with one another in this sector. 

\begin{figure}[h]
	\begin{center}
		\begin{tikzpicture}
			\draw[-{Latex},thick] (0,0) -- (5,0);
			\foreach \x in  {0,2.5,3.7} \draw[shift={(\x,0)},color=black, thick] (0pt,3pt) -- (0pt,-3pt);
			\foreach \x in {0} \draw[shift={(\x,0)},color=black] (0pt,0pt) -- (0pt,-3pt) node[below] {$\x$};
			\filldraw (4.9,-0.15) node[align=left,   below] {$\omega$};
			\filldraw (2.5,-0.15) node[align=left,   below] {$\omega_g$};
			\filldraw (3.7,-0.15) node[align=left, below] {$\omega_c$};
			\draw[red,thick] (0.4,0.4)--(1.9,0.4);
			\draw[red,thick] (2.49,0).. controls +(up:0.45cm) and +(left:0cm)..
			node[above,sloped] {} (1.9,0.4);
			\draw[red,thick] (0.015,0).. controls +(up:0.45cm) and +(left:0cm)..
			node[above,sloped] {} (0.4,0.4);
			\draw[blue,thick] (0.7,0.7)--(1.8,0.7);
			\draw[blue,thick] (2.51,0).. controls +(up:0.75cm) and +(left:0cm)..
			node[above,sloped] {} (1.8,0.7);
			\draw[blue,thick] (-0.01,0).. controls +(up:0.75cm) and +(left:0cm)..
			node[above,sloped] {} (0.7,0.7);
		\end{tikzpicture}
	\end{center}
	\caption{Intersection intervals of GTDS (red curve) and LTDS (blue curve) for fixed $(\Omega, P,\Psi)$. }
\end{figure}
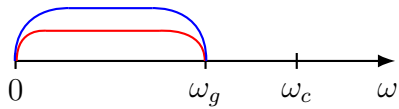

The three limiting cases are:
\begin{equation}\label{key}
C_{\Omega, P, \Psi}\mathop{=}_{\alpha\to 0^{+}} \frac{\pi ^2 T}{2 \left(4 \pi  P-\Omega ^2\right)},
\quad
C_{\Omega\to 0, P, \Psi}=\frac{\pi  T}{8 P},
\quad
C_{\Omega, P\to 0, \Psi}=-\frac{\pi ^2 T}{2 \Omega ^2}.
\end{equation}

\subsubsection{Specific heat $C_{J, P, \Psi}$}

For constant $(J, P, \Psi)$ the relevant specific heat is  $C_{J, P, \Psi}$. It
has no apparent divergences and  is always positive in this sector (see
Fig.~\ref{figHeat1}).
The three limiting cases for $C_{J, P, \Psi}$ are:
\begin{equation}\label{key}
C_{J, P, \Psi}\mathop{=}_{\alpha\to 0^{+}}\frac{2 \pi ^2 T}{16 \pi  P+5 \Omega ^2},
\quad
C_{J, P, \Psi}\mathop{=}_{\Omega\to 0}\frac{\pi  T}{8 P},
\quad
C_{J, P\to 0, \Psi}=\frac{2 \pi ^2 T}{5 \Omega ^2}.
\end{equation}

\subsubsection{Specific heat  $C_{\Omega, V, \alpha}$}

The specific heat  $C_{\Omega, V, \alpha} <0$
is always negative in this sector, thus the RGB$_3$  black hole is locally
unstable from thermodynamic standpoint with respect to fixed $(\Omega, V,
\alpha)$. There are no physical divergences occurring for this specific heat.
The three limiting cases are:
\begin{equation}\label{key}
C_{\Omega, V, \alpha\to 0^{+}}=0,
\quad
C_{\Omega\to 0, V, \alpha}=0,
\quad
C_{\Omega, V, \alpha}\mathop{=}_{P\to 0}-\frac{\pi ^2 \alpha ^2 T \Omega ^2}{2 \alpha ^2 \Omega ^4-2 \alpha  \Omega ^2+1}.
\end{equation}
We note that the limiting cases at $\alpha\to 0^+$ and $\omega\to 0$ the
specific heat $C_{\Omega, V, \alpha}$ vanishes. This is not unexpected
situation, because zero specific heat corresponds to a phase transition of the
system, which can occur when some of the parameters are taken to their limits.
This type of situations correspond to different gravitational solutions with different thermodynamics from that of the RGB$_3$ black hole.  

\subsubsection{Specific heat  $C_{\Omega, V, \Psi}$}
The denominator of $C_{\Omega, V,\Psi}$ is a quadratic function of $\omega$ with roots $\omega_-=\omega_c$ and $\omega_+$ given by
\begin{equation}\label{key}
	\omega_+ = \frac{2 (Y+1) (3 Y+4)}{2 Y^2+2 Y-1}\omega_g.
\end{equation}

 Inspecting the coefficient in front of $\omega^2$ one notices three cases, namely:
\begin{itemize}
	\item $0<\alpha <\dfrac{\sqrt{3}}{64\pi P}$.
In this sector, the specific heat is positive for
	\begin{equation}\label{key}
		0 < \omega < \omega_c,
	\end{equation}
which defines the LTDS for this case. Since $\omega_g<\omega_c$ there is an
intersection between the local and global thermodynamic stability (see
Fig.~\ref{figHeat1}).

\item $\alpha =\dfrac{\sqrt{3}}{64\pi P}$.
In this case, the LTDS region is 
\begin{equation}\label{key}
	0<\omega<\omega_c, \quad \text{where} \quad \omega_c= \left(1-\frac{\sqrt{3}}{3}\right)16\pi P.
\end{equation}
Substituting the value for $\alpha$ in $\omega_g$, we find that
$\omega<\omega_g<\omega_c$ and thus LTDS includes the GTDS\@. This is the
same situation as depicted on Fig.~\ref{figHeat1}.

\item $\alpha >\dfrac{\sqrt{3}}{64\pi P}$. In this case the LTDS is 
\begin{equation}\label{eqComegavpsiineq}
	0<\omega<\omega_c.
\end{equation}
Here $\omega_g<\omega_c$ and the situation resembles again the one depicted on 
Fig.~\ref{figHeat1}.  
\end{itemize}

Finally, the three limiting cases for  $C_{\Omega, V, \Psi}$ are:
\begin{equation}\label{key}%
C_{\Omega, V, \Psi}\mathop{=}_{\alpha\to 0^{+}}\frac{\pi ^2 T \Omega ^2}{\left(8 \pi  P-\Omega ^2\right) \left(32 \pi  P+\Omega ^2\right)},
\quad C_{\Omega\to 0, V, \Psi}=0,\quad
C_{\Omega, V, \Psi}\mathop{=}_{P\to 0}-\frac{\pi ^2 T}{\Omega ^2}.
\end{equation}
\subsubsection{Specific heat  $C_{J, V, \alpha}$}

The denominator of $C_{J, V, \alpha}$ is again a quadratic polynomial $f(\omega)$ with respect to $\omega$.
The only positive root of $f(\omega)$ is
\begin{equation}\label{key}
\omega_{+}= \frac{Y (5 Y+12)+8+\sqrt{(3 Y+4) (83 Y^3+260 Y^2+272 Y+96)}}{4 \alpha  (7 Y+10)}.
\end{equation}
One can check that the LTDS condition $C_{J, V, \alpha}>0$ requires
$f(\omega)>0$. The latter leads to  $\omega_c<\omega_+<\omega$, thus the scope
of LTDS is beyond the physical interval $0<\omega<\omega_c$. Therefore, the
RGB$_3$ black hole can not be locally nor globally stable for fixed
$(J,V,\alpha)$.  

The three limiting cases for  $C_{J, V, \alpha}$ are:

\begin{equation}\label{key}
C_{J, V,\alpha\to 0^{+}}=0,
\quad C_{J , V,\alpha}\mathop{=}_{\Omega\to 0}0, \quad C_{J,
V,\alpha}\mathop{=}_{P\to 0}\frac{2 \pi ^2 \alpha ^2 T \Omega ^2}{5 \alpha ^2
\Omega ^4-2 \alpha  \Omega ^2-1}.
\end{equation}

\subsubsection{Specific heat  $C_{J, V, \Psi}$}

The final relevant specific heat in this sector is $C_{J, V, \Psi}$. It is
always negative. Thus no LTDS or GTDS exist in this case. The three limiting
cases for $C_{J, V, \Psi}$ are:
\begin{equation}\label{key}
C_{J, V, \Psi}\mathop{=}_{\alpha\to 0^{+}}\frac{2 \pi ^2 T}{16 \pi  P+5 \Omega ^2},
\quad
C_{J, V, \Psi}\mathop{=}_{\Omega\to 0}\frac{\pi  T}{8 P},
\quad
C_{J,V, \Psi}\mathop{=}_{P\to 0}\frac{2 \pi ^2 T}{5 \Omega ^2}.
\end{equation}
However one notes that in the limiting cases, shown above, the heat capacity
$C_{J, V, \Psi}$ is positive. Once again, this is due to the fact that these
cases correspond to different than RGB$_3$ gravitational systems.
\\

Let us make a short summary of the result from this section. We have analyzed
the behavior of the physical specific heats of the RGB$_3$ black hole in
 $\alpha>0$ sector. Six of the specific heats can be positive in some regions
of the equilibrium space. In two cases, namely for fixed $(\Omega, V, \alpha)$
and fixed $(J, V, \alpha)$,  we do not have a local
thermodynamic stability. This leads to the
conclusion that there doesn't exist a sector in the phase space, where the
system is in local equilibrium with respect to all of its parameters.

\subsection{Negative Gauss–Bonnet parameter, $\alpha_p<\alpha <0$}

\subsubsection{Specific heat $C_{\Omega, P,\alpha}$}

\begin{figure}[H] 
	\begin{center}
		\begin{subfigure}{0.3\textwidth}
			\begin{tikzpicture}
				\draw[-{Latex},thick] (0,0) -- (3.7,0);
				\foreach \x in  {0,2,2.8} \draw[shift={(\x,0)},color=black, thick] (0pt,3pt) -- (0pt,-3pt);
				\foreach \x in {0} \draw[shift={(\x,0)},color=black] (0pt,0pt) -- (0pt,-3pt) node[below] {$\x$};
				\filldraw (2,-0.15) node[align=left,   below] {$\omega_g$};
				\filldraw (2.8,-0.15) node[align=left, below] {$\omega_c$};
				\filldraw (3.7,-0.15) node[align=left, below] {$\omega$};
				\draw[red,thick] (0.01,0).. controls +(up:0.43cm) and +(left:0cm)..
				node[above,sloped] {} (0.4,0.4);
				\draw[red,thick] (2,0.4)--(2,0);
				\draw[red,thick] (0.4,0.4)--(2.01,0.4);
				\draw[blue,thick] (-0.01,0).. controls +(up:0.75cm) and +(left:0cm)..
				node[above,sloped] {} (0.699,0.7);
				\draw[blue,thick] (2.8,0).. controls +(up:0.75cm) and +(left:0cm)..
				node[above,sloped] {} (2.1,0.7);
				\draw[blue,thick] (0.7,0.7)--(2.11,0.7);
			\end{tikzpicture}
			\caption{$-\frac{1}{36\pi P}\leq \alpha< 0$.}\label{FigGTDSAneg1a}
		\end{subfigure}
		\begin{subfigure}{0.3\textwidth}
			\begin{tikzpicture}
				\draw[-{Latex},thick] (0,0) -- (4.5,0);
				\foreach \x in  {0,1.2,2,3,3.7} \draw[shift={(\x,0)},color=black, thick] (0pt,3pt) -- (0pt,-3pt);
				\foreach \x in {0} \draw[shift={(\x,0)},color=black] (0pt,0pt) -- (0pt,-3pt) node[below] {$\x$};
				\filldraw (1.2,-0.15) node[align=left,   below] {$\omega_+$};
				\filldraw (2,-0.15) node[align=left,   below] {$\omega_-$};
				\filldraw (3,-0.15) node[align=left,   below] {$\omega_g$};
				\filldraw (3.7,-0.15) node[align=left, below] {$\omega_c$};
				\filldraw (4.5,-0.15) node[align=left, below] {$\omega$};
				\draw[red,thick] (0.01,0).. controls +(up:0.45cm) and +(left:0cm)..
				node[above,sloped] {} (0.4,0.4);
				\draw[red,thick] (0.4,0.4)--(1.21,0.4);
				\draw[red,thick] (1.2,0.4)--(1.2,0);
				\draw[red,thick] (2,0)--(2,0.4);
				\draw[red,thick] (1.99,0.4)--(3,0.4);
				\draw[red,thick] (3,0.4)--(3,0);
				\draw[blue,thick] (-0.01,0).. controls +(up:0.73cm) and +(left:0cm)..
				node[above,sloped] {} (0.699,0.7);
				\draw[blue,thick] (3.7,0).. controls +(up:0.73cm) and +(left:0cm)..
				node[above,sloped] {} (3,0.7);
				\draw[blue,thick] (0.699,0.7)--(3,0.7);
			\end{tikzpicture}
			\caption{$\alpha_p< \alpha\leq -\frac{3}{100\pi P}$.}\label{FigGTDSAneg2b}
		\end{subfigure}
	\end{center}
	\vspace{-10pt}
  \caption{Intervals of thermodynamic stability:  a) GTDS for $-1/(36\pi P)\leq\alpha< 0$
  occurs in the interval $0< \omega\leq\omega_g$ (red curve) and LTDS occurs in the interval given by
$0<\omega<\omega_c$ (blue curve). b) GTDS for $\alpha_p<\alpha\leq-3/(100\pi
P)$ occurs within $0<\omega\leq\omega_{+}$ or $\omega_{-}\leq \omega\leq\omega_{g}$
(red curves) , and the LTDS occurs when $0<\omega<\omega_c$ (blue
curve).}\label{FigGTDS1}
\end{figure}
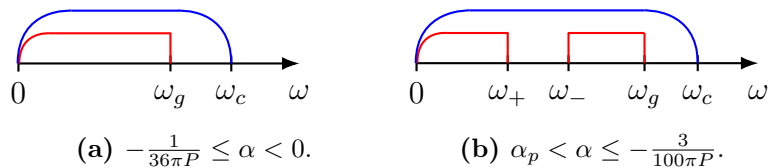

In this case the condition for LTDS $(C_{\Omega, P,\alpha}>0)$ reduces to
\begin{equation}\label{key}
	0<\omega<\omega_c.
\end{equation} 
Now let us check if the global and the local thermodynamic stability intersect. There are three relevant cases as depicted on Fig.~\ref{FigGTDS1}.

\subsubsection{Specific heat $C_{J, P,\alpha}$}

In this case  $C_{J, P,\alpha}$ is always positive and the black hole is in
LTDS  for all $\omega<\omega_c$. This situation corresponds to Fig.~\ref{FigGTDS1}.

\subsubsection{Specific heat $C_{\Omega, P, \Psi}$}
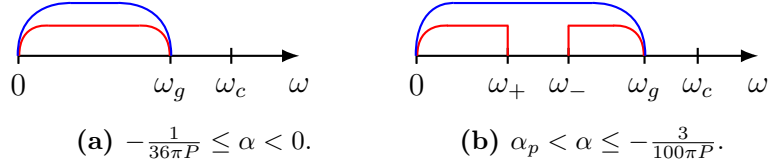
\begin{figure}[h] 
	\begin{center}
		\begin{subfigure}{0.3\textwidth}
			\begin{tikzpicture}
				\draw[-{Latex},thick] (0,0) -- (3.7,0);
				\foreach \x in  {0,2,2.8} \draw[shift={(\x,0)},color=black, thick] (0pt,3pt) -- (0pt,-3pt);
				\foreach \x in {0} \draw[shift={(\x,0)},color=black] (0pt,0pt) -- (0pt,-3pt) node[below] {$\x$};
				\filldraw (2,-0.15) node[align=left,   below] {$\omega_g$};
				\filldraw (2.8,-0.15) node[align=left, below] {$\omega_c$};
				\filldraw (3.7,-0.15) node[align=left, below] {$\omega$};
				\draw[red,thick] (0.01,0).. controls +(up:0.45cm) and +(left:0cm)..
				node[above,sloped] {} (0.39,0.4);
				\draw[red,thick] (1.99,0).. controls +(up:0.45cm) and +(left:0cm)..
				node[above,sloped] {} (1.59,0.4);
				\draw[red,thick] (0.39,0.4)--(1.59,0.4);
				\draw[blue,thick] (-0.01,0).. controls +(up:0.75cm) and +(left:0cm)..
				node[above,sloped] {} (0.69,0.7);
				\draw[blue,thick] (2.01,0).. controls +(up:0.75cm) and +(left:0cm)..
				node[above,sloped] {} (1.31,0.7);
				\draw[blue,thick] (0.69,0.7)--(1.31,0.7);
				%\draw [out=90,in=90,color=red,thick]  (0,0) to (2.2,0);
			\end{tikzpicture}
			\caption{$-\frac{1}{36\pi P}\leq \alpha< 0$.}\label{FigGTDSAneg1a2}
		\end{subfigure}
		\begin{subfigure}{0.3\textwidth}
			\begin{tikzpicture}
				\draw[-{Latex},thick] (0,0) -- (4.5,0);
				\foreach \x in  {0,1.2,2,3,3.7} \draw[shift={(\x,0)},color=black, thick] (0pt,3pt) -- (0pt,-3pt);
				\foreach \x in {0} \draw[shift={(\x,0)},color=black] (0pt,0pt) -- (0pt,-3pt) node[below] {$\x$};
				\filldraw (1.2,-0.15) node[align=left,   below] {$\omega_+$};
				\filldraw (2,-0.15) node[align=left,   below] {$\omega_-$};
				\filldraw (3,-0.15) node[align=left,   below] {$\omega_g$};
				\filldraw (3.7,-0.15) node[align=left, below] {$\omega_c$};
				\filldraw (4.5,-0.15) node[align=left, below] {$\omega$};
				\draw[red,thick] (0.01,0).. controls +(up:0.45cm) and +(left:0cm)..
				node[above,sloped] {} (0.41,0.4);
				\draw[red,thick] (2.99,0).. controls +(up:0.45cm) and +(left:0cm)..
				node[above,sloped] {} (2.59,0.4);
				\draw[red,thick] (0.4,0.4)--(1.21,0.4);
				\draw[red,thick] (1.2,0.4)--(1.2,0);
				\draw[red,thick] (2,0)--(2,0.4);
				\draw[red,thick] (1.99,0.4)--(2.6,0.4);
				
				\draw[blue,thick] (-0.01,0).. controls +(up:0.75cm) and +(left:0cm)..
				node[above,sloped] {} (0.49,0.7);
				\draw[blue,thick] (3.01,0).. controls +(up:0.75cm) and +(left:0cm)..
				node[above,sloped] {} (2.31,0.7);
				\draw[blue,thick] (0.49,0.7)--(2.31,0.7);
				%\draw [out=90,in=90,color=red,thick]  (0,0) to (2.2,0);
			\end{tikzpicture}
			\caption{$\alpha_p< \alpha\leq -\frac{3}{100\pi P}$.}\label{FigGTDSAneg2b2}
		\end{subfigure}
	\end{center}
	\vspace{-10pt}
  \caption{Intervals of thermodynamic stability:  a) GTDS for $-1/(36\pi P)\leq\alpha< 0$
  occurs within $0< \omega<\omega_g$ (red curve) and LTDS occurs in the interval
$0<\omega<\omega_g$ (blue curve). b) GTDS for $\alpha_p<\alpha\leq-3/(100\pi
P)$ occurs when $0<\omega\leq\omega_{+}$ or $\omega_{-}\leq \omega<\omega_{g}$
(red curves), and the LTDS is defined by $0< \omega<\omega_g$ (blue curve).}
\label{FigGTDS2}
\end{figure}

The LTDS in this case is given by 
\begin{equation}\label{key}
	0<\omega<\omega_g.
\end{equation}
The comparison with GTDS is shown on Fig.~\ref{FigGTDS2}.

\subsubsection{Specific heat $C_{J, P, \Psi}$}
The specific heat $C_{J, P,\Psi}$ is always positive and the black hole is in LTDS
for all $0<\omega<\omega_c$. This situation corresponds to Fig.~\ref{FigGTDS1}.

\subsubsection{Specific heat  $C_{\Omega, V, \alpha}$}
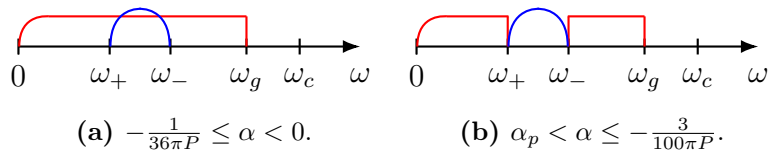
\begin{figure}[h] 
	\begin{center}
		\begin{subfigure}{0.3\textwidth}
			\begin{tikzpicture}
				\draw[-{Latex},thick] (0,0) -- (4.5,0);
				\foreach \x in  {0,1.2,2,3,3.7} \draw[shift={(\x,0)},color=black, thick] (0pt,3pt) -- (0pt,-3pt);
				\foreach \x in {0} \draw[shift={(\x,0)},color=black] (0pt,0pt) -- (0pt,-3pt) node[below] {$\x$};
				\filldraw (1.2,-0.15) node[align=left,   below] {$\omega_+$};
				\filldraw (2,-0.15) node[align=left,   below] {$\omega_-$};
				\filldraw (3,-0.15) node[align=left,   below] {$\omega_g$};
				\filldraw (3.7,-0.15) node[align=left, below] {$\omega_c$};
				\filldraw (4.5,-0.15) node[align=left, below] {$\omega$};
				\draw[red,thick] (0.01,0).. controls +(up:0.45cm) and +(left:0cm)..
				node[above,sloped] {} (0.4,0.4);
				\draw[red,thick] (0.4,0.4)--(3,0.4);
				\draw[red,thick] (3,0.4)--(3,0);
				\draw[blue,thick] (1.21,0).. controls +(up:0.53cm) and +(left:0cm)..
				node[above,sloped] {} (1.61,0.5);
				\draw[blue,thick] (1.99,0).. controls +(up:0.53cm) and +(left:0cm)..
				node[above,sloped] {} (1.59,0.5);
			\end{tikzpicture}
			\caption{$-\frac{1}{36\pi P}\leq \alpha< 0$.}\label{FigGTDSAneg1a21}
		\end{subfigure}
		\begin{subfigure}{0.3\textwidth}
			\begin{tikzpicture}
				\draw[-{Latex},thick] (0,0) -- (4.5,0);
				\foreach \x in  {0,1.2,2,3,3.7} \draw[shift={(\x,0)},color=black, thick] (0pt,3pt) -- (0pt,-3pt);
				\foreach \x in {0} \draw[shift={(\x,0)},color=black] (0pt,0pt) -- (0pt,-3pt) node[below] {$\x$};
				\filldraw (1.2,-0.15) node[align=left,   below] {$\omega_+$};
				\filldraw (2,-0.15) node[align=left,   below] {$\omega_-$};
				\filldraw (3,-0.15) node[align=left,   below] {$\omega_g$};
				\filldraw (3.7,-0.15) node[align=left, below] {$\omega_c$};
				\filldraw (4.5,-0.15) node[align=left, below] {$\omega$};
				\draw[red,thick] (0.01,0).. controls +(up:0.45cm) and +(left:0cm)..
				node[above,sloped] {} (0.4,0.4);
				\draw[red,thick] (0.4,0.4)--(1.21,0.4);
				\draw[red,thick] (1.2,0.4)--(1.2,0);
				\draw[red,thick] (2,0)--(2,0.4);
				\draw[red,thick] (1.99,0.4)--(3,0.4);
				\draw[red,thick] (3,0.4)--(3,0);
				\draw[blue,thick] (1.21,0).. controls +(up:0.53cm) and +(left:0cm)..
				node[above,sloped] {} (1.61,0.5);
				\draw[blue,thick] (1.99,0).. controls +(up:0.53cm) and +(left:0cm)..
				node[above,sloped] {} (1.59,0.5);
			\end{tikzpicture}
			\caption{$\alpha_p< \alpha\leq -\frac{3}{100\pi P}$.}\label{FigGTDSAneg2b13}
		\end{subfigure}
	\end{center}
	\vspace{-10pt}
  \caption{Intervals of thermodynamic stability: a) GTDS for $-1/(36\pi P)\leq\alpha<0$
  occurs when $0< \omega\leq\omega_g$ (red curve) and LTDS occurs in the interval
$\omega_+<\omega<\omega_-$ (blue curve). b) GTDS for
$\alpha_p<\alpha\leq-3/(100\pi P)$ occurs in the intervals $0<\omega\leq\omega_{+}$ or
$\omega_{-}\leq \omega\leq\omega_{g}$ (red curves), and the LTDS occurs for
$\omega_+<\omega<\omega_-$ (blue curve).}\label{FigGTDS134}
\end{figure}

In the positive $\alpha>0$ case this specific heat was always negative. It turns
out that this is not the case for $\alpha<0$. For $\alpha \leq -\dfrac{3}{100\pi
P}$ both roots of its denominator are positive and so $C_{\Omega,
V,\alpha}>0$, thus the LTDS is given by 
\begin{equation}
\omega_+<\omega<\omega_-, \quad \alpha \leq -\dfrac{3}{100\pi
	P},
\end{equation}
where $\omega_\pm$ are defined in \eqref{key1}. This case is illustrated in
Fig.~\ref{FigGTDS134}.

\subsubsection{Specific heat  $C_{\Omega, V, \Psi}$}
For this specific heat one can show that the LTDS condition is
$0<\omega<\omega_c$. Furthermore, the intersection between LTDS and GTDS is again depicted by Fig.~\ref{FigGTDS1}.

\subsubsection{Specific heat  $C_{J, V, \alpha}$}
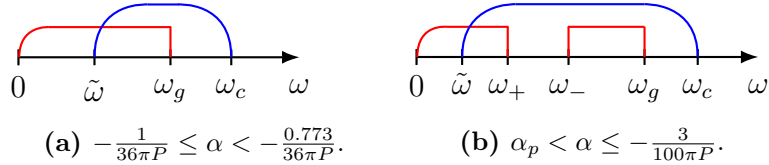
\begin{figure}[h] 
	\begin{center}
		\begin{subfigure}{0.3\textwidth}
			\begin{tikzpicture}
				\draw[-{Latex},thick] (0,0) -- (3.7,0);
				\foreach \x in  {0,1,2,2.8} \draw[shift={(\x,0)},color=black, thick] (0pt,3pt) -- (0pt,-3pt);
				\foreach \x in {0} \draw[shift={(\x,0)},color=black] (0pt,0pt) -- (0pt,-3pt) node[below] {$\x$};
				\filldraw (1,-0.15) node[align=left,   below] {$\tilde\omega$};
				\filldraw (2,-0.15) node[align=left,   below] {$\omega_g$};
				\filldraw (2.8,-0.15) node[align=left, below] {$\omega_c$};
				\filldraw (3.7,-0.15) node[align=left, below] {$\omega$};
				\draw[red,thick] (0.01,0).. controls +(up:0.43cm) and +(left:0cm)..
				node[above,sloped] {} (0.4,0.4);
				\draw[red,thick] (2,0.4)--(2,0);
				\draw[red,thick] (0.4,0.4)--(2.01,0.4);
				\draw[blue,thick] (1,0).. controls +(up:0.75cm) and +(left:0cm)..
				node[above,sloped] {} (1.7,0.7);
				\draw[blue,thick] (2.8,0).. controls +(up:0.75cm) and +(left:0cm)..
				node[above,sloped] {} (2.1,0.7);
				\draw[blue,thick] (1.7,0.7)--(2.11,0.7);
			\end{tikzpicture}
			\caption{$-\frac{1}{36\pi P}\leq \alpha< -\frac{0.773}{36\pi P}$.}\label{FigGTDSAneg1a212}
		\end{subfigure}
		\begin{subfigure}{0.3\textwidth}
			\begin{tikzpicture}
				\draw[-{Latex},thick] (0,0) -- (4.5,0);
				\foreach \x in  {0,0.6,1.2,2,3,3.7} \draw[shift={(\x,0)},color=black, thick] (0pt,3pt) -- (0pt,-3pt);
				\foreach \x in {0} \draw[shift={(\x,0)},color=black] (0pt,0pt) -- (0pt,-3pt) node[below] {$\x$};
				\filldraw (0.6,-0.07) node[align=left,   below] {$\tilde\omega$};
				\filldraw (1.2,-0.15) node[align=left,   below] {$\omega_+$};
				\filldraw (2,-0.15) node[align=left,   below] {$\omega_-$};
				\filldraw (3,-0.15) node[align=left,   below] {$\omega_g$};
				\filldraw (3.7,-0.15) node[align=left, below] {$\omega_c$};
				\filldraw (4.5,-0.15) node[align=left, below] {$\omega$};
				\draw[red,thick] (0.01,0).. controls +(up:0.45cm) and +(left:0cm)..
				node[above,sloped] {} (0.4,0.4);
				\draw[red,thick] (0.4,0.4)--(1.21,0.4);
				\draw[red,thick] (1.2,0.4)--(1.2,0);
				\draw[red,thick] (2,0)--(2,0.4);
				\draw[red,thick] (1.99,0.4)--(3,0.4);
				\draw[red,thick] (3,0.4)--(3,0);
				\draw[blue,thick] (0.6,0).. controls +(up:0.73cm) and +(left:0cm)..
				node[above,sloped] {} (01.3,0.7);
				\draw[blue,thick] (3.7,0).. controls +(up:0.73cm) and +(left:0cm)..
				node[above,sloped] {} (3,0.7);
				\draw[blue,thick] (1.3,0.7)--(3,0.7);
			\end{tikzpicture}
			\caption{$\alpha_p< \alpha\leq -\frac{3}{100\pi P}$.}\label{FigGTDSAneg2b123}
		\end{subfigure}
	\end{center}
	\vspace{-10pt}
  \caption{Intervals of thermodynamic stability:  a) GTDS for $-1/(36\pi P)\leq\alpha<
  -\frac{0.773}{36\pi P}$ occurs in the interval $\tilde\omega< \omega\leq\omega_g$ (red
curve) and LTDS occurs for $\tilde\omega<\omega<\omega_c$ (blue curve). b) GTDS
for $\alpha_p<\alpha\leq-3/(100\pi P)$ occurs within
$\tilde\omega<\omega\leq\omega_{+}$ or $\omega_{-}\leq \omega\leq\omega_{g}$
(red curves), and the LTDS occurs when $\tilde\omega<\omega<\omega_c$ (blue
curve).}\label{FigGTDS1344}
\end{figure}

The denominator of this specific heat has only one positive root,
\begin{equation}\label{key}
\tilde\omega=\frac{Y (5 Y+12)+8-\sqrt{(3 Y+4) (83Y^3+260Y^2+272Y+96)}}{4 \alpha  (7 Y+10)}.
\end{equation}
The LTDS condition in this case
is satisfied by $\tilde\omega<\omega<\omega_c$ and $\alpha<-\frac{5}{288\pi P}$. The first GTDS case $(-1/36\pi P<\alpha<0)$ is further bounded by 
\begin{equation}
	\alpha < -\frac{1-x_2}{32\pi P}\approx -\frac{0.773}{36\pi P},
\end{equation}
where $x_2 \approx 0.313$. This point comes from the intersection of the curves
$\tilde \omega=\omega_g$, resulting in  the polynomial equation
\begin{equation}
	4 + 9 x - 2 x^2 - 24 x^3 - 2 x^4 + 7 x^5=0.
\end{equation}
Therefore, below the intersection point $x_2$ one has
$\tilde\omega<\omega\leq\omega_g<\omega_c$, hence LTDS and GTDS have a common
intersection region as shown on Fig.~\ref{FigGTDSAneg1a212}. 

For the second GTDS case \eqref{eqGTDS2} more complex situation is realized. It is given by two non-intersecting intervals:
\begin{equation}\label{key}
\tilde\omega<\omega\leq\omega_+, \quad \omega_-\leq\omega\leq\omega_g.
\end{equation}
The occurrence is depicted on Fig.~\ref{FigGTDSAneg2b123}.
\subsubsection{Specific heat  $C_{J, V, \Psi}$}

The specific heat  $C_{J, V, \Psi}$ is always positive, thus the RGB$_3$ black hole is locally stable for $0<\omega<\omega_c$.
The comparison with GTDS is depicted by Fig.~\ref{FigGTDS1}.
\\

As a short summary: we found that for $\alpha<0$ all heat capacities acquire regions of local thermodynamic stability. 
Contrary to the situation in the previous sector for $\alpha>0$, now one could find a region,
where the black hole is thermodynamically stable in all of its parameters.

\section{Conclusion}\label{secConcl}
In the present paper we have analyzed the conditions for local and global thermodynamic equilibrium of the 3-dimensional rotating
Gauss-Bonnet black holes. We have presented a full analysis of the global
thermodynamic stability in the weak global conjecture, utilizing the most natural thermodynamic potential for
the given ensemble of macro parameters.
Since all of the state quantities in this ensemble share
a common divergence it turns out that physical states occur for values of the
angular velocity $\Omega^2 < (\sqrt{32\pi\alpha P +1}-1)/(2\alpha)$. Our study included the weak global thermodynamic stability in both sectors 
of the Gauss–Bonnet parameter $\alpha$. 

Due to the fact that global thermodynamic stability implies local one, we have performed an exhaustive analysis of the local
thermodynamic picture. This was done via the Nambu bracket formalism, developed
in \cite{Mansoori:2014oia}. All of the 8 possible specific heats have been
analyzed for both sectors for $\alpha$. Interestingly, in the $\alpha>0$ case,
not all heat capacities admit a region of local thermodynamic stability.
Namely, the specific heats $C_{\Omega,V,\alpha}$ and $C_{J,V,\Psi}$ are always
negative. The missing underlying local stability also implies that for fixed $({\Omega,V,\alpha})$ and $({J,V,\Psi})$ a global one can not be established. For
$\alpha_p<\alpha<0$, this is not the case as all specific heats have
a region of positivity and thus the black hole can be locally
stable from thermodynamic standpoint.

It is
natural to assume that true thermodynamic equilibrium can only be properly
established in regions where local and global thermodynamic stability occur at
the same time. For this reason, we have looked for intersections between LTDS
and GTDS in both sectors for the Gauss-Bonnet parameter $\alpha$. We have discovered
that proper equilibrium exists for all specific heats in the case
$\alpha_p<\alpha<0$. In the $\alpha>0$ this is true only with respect to some
of the specific heats. When LTDS and GTDS intersect non-trivial conditions on
$\alpha$ as a function of $P$ emerge, which highly restricts the physics in
this regions.

To our surprise, the global thermodynamic analysis and its relations to the
local one till now has not been presented in full for black holes. For this
reason we felt compelled to state it clearly for the first time. Although we
presented it for the energy potential in its fullness on a three dimensional gravitational system, it holds valid in any dimensions,
whenever there is a well-defined first law of thermodynamics. However, in general it is still not known how to define the conditions for the strong global thermodynamic stability, when passing to a different potential. In this case, the Legendre transformation allows one to define correctly only the weak global stability conditions on the new potential.

This paper is intended to be the first  of series of papers, where  different
aspects of the RGB$_3$ black hole will be investigated. One direction is to
consider the thermodynamic geometry, where one can study the proper
thermodynamic metrics on the space of the equilibrium states of the black hole.
Investigating the holographic
complexity of the RGB$_3$ black hole is another interesting problem. Studying
the role of non-extensive thermodynamics over the extensive one presents yet
another challenge. As mentioned previously, finding the conditions for the strong global thermodynamic stability, when passing to a different potential, is also very challenging. Finally, one can extend this work by including
non-perturbative correction to the entropy, where the  new coupling parameters
in the correction terms can be constrained in a highly non-trivial way. It must
be noted that the type of analysis presented in the current work can also be
applied to a broad class of multi-parameter thermal systems besides black
holes. 

\section*{Acknowledgments}

The authors would like to thank Dimitar Marvakov for the invaluable comments on
both local and global aspects of thermodynamics. T. V. is also grateful to  Seyed
Ali Hosseini Mansoori for  comments on specific heats and the local
thermodynamic stability. I. I. and M. R. gratefully
acknowledge the support of the Bulgarian national program ``Young Scientists
and Postdoctoral Research Fellows''. This work
was partially supported by the Bulgarian NSF grant N28/5 as well as the program
``JINR - Bulgaria'' of the Bulgarian Nuclear Regulatory Agency.

\appendix

\section{Weak GTDS conditions for other thermodynamic potentials}\label{appAB}

In order to make our analysis more complete, let us say few words about
some of the other thermodynamic potentials. Due to the fact that different
potentials correspond to different constraints to which the system may be
subjected one can study GTDS  by constructing other energy derived
thermodynamic potentials (see Appendix~\ref{appA}). The latter can be obtained
by the proper Legendre transformation of the energy potential along given
natural state quantities. For example the enthalpy of spacetime, the Gibbs free
energy and the Helmholtz free energy are given by
\begin{align}\label{key}
	&M=H=\mathcal{L}_V E=E-(-P V)=E+ P V,
	\\
	&G=\mathcal{L}_{S,V} E=E-TS+P V,
	\\
	&F=\mathcal{L}_S E=E-T S.
\end{align}
The natural parameters of these potentials can be obtained from the corresponding form of the first law:
\begin{align}\label{key}
	&	\delta M=T\delta S+ \Omega \delta J+V\delta P+\Psi\delta \alpha,
	\\
	&	\delta G=-S\delta T+ \Omega \delta J+V\delta P+\Psi\delta \alpha,
	\\
	&	\delta F=-S\delta T+ \Omega \delta J-P\delta V+\Psi\delta \alpha,
\end{align}
which lead to $M=M(S, J, P, \alpha)$, $G=G(T, J, P, \alpha)$ and $F=F(T, J, V, \alpha)$. 

Let us now consider the conditions for weak global thermodynamic stability. For
example, when considering the mass potential $M$ the Legendre transformation  of $E$ along
$V$ preserves the sign of the inequalities from (\ref{eqGlobalCondE})
except for the conjugate $P$ of $V$. Thus the weak conditions for the
minimum of the mass potential in equilibrium now read:
\begin{equation}\label{key}
	\frac{\partial^2 M}{\partial S^2}\Big |_{J, P, \alpha}\geq 0,
	\quad \frac{\partial^2 M}{\partial J^2}\Big |_{S, P, \alpha}\geq 0,
	\quad \frac{\partial^2 M}{\partial P^2}\Big |_{S, J, \alpha}\leq 0,
	\quad \frac{\partial^2 M}{\partial \alpha^2}\Big |_{S,J,P }\geq 0.
\end{equation}
Therefore the mass is a convex function of $S, J$ and $\alpha$, but a concave function of $P$.
Similar reasoning holds for $G$ and $F$:
\begin{equation}\label{key}
	\frac{\partial^2 G}{\partial T^2}\Big |_{J, P, \alpha}\leq 0,
	\quad \frac{\partial^2 G}{\partial J^2}\Big |_{T, P, \alpha}\geq 0,
	\quad \frac{\partial^2 G}{\partial P^2}\Big |_{T, J, \alpha}\leq 0,
	\quad \frac{\partial^2 G}{\partial \alpha^2}\Big |_{T,J,P }\geq 0.
\end{equation}
\begin{equation}\label{key}
	\frac{\partial^2 F}{\partial T^2}\Big |_{J, V, \alpha}\leq 0,
	\quad \frac{\partial^2 F}{\partial J^2}\Big |_{T, V, \alpha}\geq 0,
	\quad \frac{\partial^2 F}{\partial V^2}\Big |_{T, J, \alpha}\geq 0,
	\quad \frac{\partial^2 F}{\partial \alpha^2}\Big |_{T,J,V }\geq 0.
\end{equation}
Therefore, the Gibbs potential is convex a function of $J$ and $ \alpha$, but
a concave function along $T$ and $P$. The Helmholtz potential is convex in $J, V$ and
$\alpha$, but concave along $T$.

Using the Legendre transformation of the energy $E=E(S, J, V, \alpha)$ one can
construct more energy derived thermodynamic potentials for the RGB$_3$ black hole. The full list
is given in Appendix~\ref{appA1}.

The energy derived thermodynamic potentials are not the only possibility. For
example, if one starts with the entropy potential one can use the Legendre
transformation of the entropy to construct new thermodynamic potentials, called
Massieu–Planck or free entropies\footnote{Sometimes they are called free
	information.}.The full list for the RGB$_3$ black hole
is given in Appendix~\ref{appA2}. To see how to do that, one rewrites the
first law with respect to the entropy
\begin{equation}\label{eqFirstLawS}
	\delta S=\frac{1}{T}\delta E-\frac{\Omega}{T} \delta J+\frac{P}{T}\delta V-\frac{\Psi}{T}\delta \alpha,
\end{equation}
where the parameter $\beta=1/T$ is the conjugate variable of $E$, the parameter
$\Omega/T$ is conjugate to $J$ and so on. Now it is obvious that the natural
parameters for the entropy are $S=S(E, J, V, \alpha)$. In equilibrium the
entropy is maximal, thus it is  globally concave in its natural parameters,
which means that its Hessian should be negative semi-definite. If one considers processes with only one fluctuating state quantity, then the weak conditions for global thermodynamic stability are
\begin{equation}\label{eqGlobalS}
	\frac{\partial^2 S}{\partial E^2}\Big |_{J, V, \alpha}\leq 0,
	\quad \frac{\partial^2 S}{\partial J^2}\Big |_{E, V, \alpha}\leq 0,
	\quad \frac{\partial^2 S}{\partial V^2}\Big |_{E, J, \alpha}\leq 0,
	\quad \frac{\partial^2 S}{\partial \alpha^2}\Big |_{E, J, V}\leq 0.
\end{equation}

The relevant Massieu–Planck potential in ($T,\Omega,P,\alpha$) space is
\begin{equation}\label{key}
	\Sigma=\mathcal{L}_{E, J, V} S=S-\frac{1}{T} E+\frac{\Omega}{T} J-\frac{P}{T} V=\frac{\pi  T Y}{16 P Y-32 \alpha  P \Omega ^2}.
\end{equation}
The first law now changes to\footnote{The natural parameters for the $\Sigma$
	potential are $(\beta, \Omega, P, \alpha)$, where $\beta=1/T$.}
\begin{equation}\label{key}
	\delta\Sigma=-(E-\Omega J+P V) \delta \frac{1}{T}+\frac{J}{T} \delta
	\Omega-\frac{V}{T} \delta P-\frac{\Psi}{T}\delta \alpha.
\end{equation}
The conditions for weak global thermodynamic stability for $\Sigma$ change sign along $T, \Omega$
and $P$ as compared to the inequalities along their conjugate variables $E, J$
and $V$ from (\ref{eqGlobalS}):
\begin{equation}\label{key}
	\frac{\partial^2 \Sigma}{\partial T^2}\Big |_{\Omega, P,\alpha}\geq 0,
	\quad \frac{\partial^2 \Sigma}{\partial \Omega^2}\Big |_{T, P, \alpha}\geq 0,
	\quad \frac{\partial^2 \Sigma}{\partial P^2}\Big |_{T,\Omega, \alpha}\geq 0,
	\quad \frac{\partial^2 \Sigma}{\partial \alpha^2}\Big |_{T, \Omega, P}\leq 0.
\end{equation}
These conditions lead to the same regions of global thermodynamic stability derived in Subsections \ref{subsecPosA} and  \ref{secGTDSneg}. This confirms the correctness of our
global thermodynamic analysis based on the Legendre transformation.

\section{Energy and entropy derived thermodynamic potentials}\label{appA}

\subsection{Energy derived thermodynamic potentials}\label{appA1}
Using the Legendre transformation of the energy $E=E(S, J, V, \alpha)$ one can
derive the following  thermodynamic potentials for the RGB$_3$ black hole:
\begin{align}\label{key}
	&\mathcal{L}_S E=E- T S,
	\\
	&\mathcal{L}_J E=E-\Omega J,
	\\
	&\mathcal{L}_V E=E+P V,
	\\
	&\mathcal{L}_\alpha E=E-\Psi \alpha,
	\\
	&\mathcal{L}_{S,J} E=E-T S-\Omega J,
	\\
	&\mathcal{L}_{S,V} E=E-T S+P V,
	\\
	&\mathcal{L}_{S,\alpha} E=E-T S-\Psi\alpha,
	\\
	&\mathcal{L}_{J,V} E= E-\Omega J+P V,
	\\
	&\mathcal{L}_{J,\alpha} E=E-\Omega J-\Psi\alpha,
	\\
	&\mathcal{L}_{V, \alpha} E=E+P V-\Psi\alpha,
	\\
	\label{eqSJVPotential}
	&\mathcal{L}_{S, J, V} E=E-T S-\Omega J+P V,
	\\
	&\mathcal{L}_{S,J,\alpha} E=E- T S-\Omega J-\Psi \alpha,
	\\
	&\mathcal{L}_{S, V, \alpha} E=E-TS +PV -\Psi \alpha,
	\\
	&\mathcal{L}_{J, V, \alpha} E=E-\Omega J+P V-\Psi \alpha,
	\\
	&\mathcal{L}_{S,J, V, \alpha} E=E-T S- \Omega J+P V-\Psi \alpha.
\end{align}
This list of thermodynamic potentials include all the standard ones (Gibbs free
energy, Helmholtz free energy, enthalpy etc.).

\subsection{Entropy derived thermodynamic potentials}\label{appA2}

Using the Legendre transformation of the entropy $S=S(E, J, V, \alpha)$ one can
derive the following Massieu–Planck thermodynamic potentials  for the RGB$_3$
black hole:
\begin{align}\label{key}
	&\mathcal{L}_E S=S- \frac{E}{T},
	\\
	&\mathcal{L}_J S=S+\frac{\Omega J}{T},
	\\
	&\mathcal{L}_V S=S-\frac{P V}{T},
	\\
	&\mathcal{L}_\alpha S=S+\frac{\Psi \alpha}{T},
	\\
	&\mathcal{L}_{E,J} S=S-\frac{E}{T}+\frac{\Omega J}{T} ,
	\\
	&\mathcal{L}_{E,V} S=S-\frac{E}{T}-\frac{PV}{T},
	\\
	&\mathcal{L}_{E,\alpha} S=S-\frac{E}{T}+\frac{\Psi\alpha}{T},
	\\
	&\mathcal{L}_{J,V} S= S+\frac{\Omega J}{T}-\frac{P V}{T},
	\\
	&\mathcal{L}_{J,\alpha} S=S+\frac{\Omega J}{T}+\frac{\Psi\alpha}{T},
	\\
	&\mathcal{L}_{V, \alpha} S=S-\frac{P V}{T}+\frac{\Psi\alpha}{T},
	\\
	\label{eqSJVPotential}
	&\mathcal{L}_{E, J, V} S=S-\frac{E}{T}+\frac{\Omega J}{T}-\frac{P V}{T},
	\\
	&\mathcal{L}_{E,J,\alpha} S=S- \frac{E}{T}+\frac{\Omega J}{T}+\frac{\Psi \alpha}{T},
	\\
	&\mathcal{L}_{E, V, \alpha} S=S-\frac{E}{T} -\frac{PV}{T} +\frac{\Psi \alpha}{T},
	\\
	&\mathcal{L}_{J, V, \alpha} S=S+\frac{\Omega J}{T}-\frac{P V}{T}+\frac{\Psi \alpha}{T},
	\\
	&\mathcal{L}_{S,J, V, \alpha} S=S-\frac{E}{T} + \frac{\Omega J}{T}-\frac{P V}{T}+\frac{\Psi \alpha}{T}.
\end{align}

This list include all the standard free entropy potentials (Gibbs free entropy,
Helmholtz free entropy, Plank potential, etc.).

\section{Nambu brackets and specific heats}\label{appB}

The local heat capacities in ($T,\Omega,P,\alpha$) space of the RGB$_3$ black hole are given by 
\begin{align}\label{key}
  &C_{J, P,\alpha}=T \left(\frac{\partial S}{\partial T}\right)_{J, P,
  \alpha}=T \frac{\{S, J, P, \alpha\}_{T, \Omega, P, \alpha}}{\{T, J, P,
\alpha\}_{T, \Omega, P, \alpha}},
	\\
  &C_{J, V,\alpha}=T \left(\frac{\partial S}{\partial T}\right)_{J, V,
  \alpha}=T \frac{\{S, J, V, \alpha\}_{T, \Omega, P, \alpha}}{\{T, J, V,
\alpha\}_{T, \Omega, P, \alpha}},
	\\
  &C_{J, P,\Psi}=T \left(\frac{\partial S}{\partial T}\right)_{J, P, \Psi}=T
  \frac{\{S, J, P, \Psi\}_{T, \Omega, P, \alpha}}{\{T, J, P, \Psi\}_{T, \Omega,
  P, \alpha}},
	\\
  &C_{J, V,\Psi}=T \left(\frac{\partial S}{\partial T}\right)_{J, V, \Psi}=T
  \frac{\{S, J, V, \Psi\}_{T, \Omega, P, \alpha}}{\{T, J, V, \Psi\}_{T, \Omega,
  P, \alpha}},
	\\
  & C_{\Omega, P, \alpha}=T \left(\frac{\partial S}{\partial T}\right)_{\Omega,
  P, \alpha}=T \frac{\{S, \Omega, P, \alpha\}_{T, \Omega, P, \alpha}}{\{T,
\Omega, P, \alpha\}_{T, \Omega, P, \alpha}},
	\\
  & C_{\Omega, V, \alpha}=T \left(\frac{\partial S}{\partial T}\right)_{\Omega,
  V, \alpha}=T \frac{\{S, \Omega, V, \alpha\}_{T, \Omega, P, \alpha}}{\{T,
\Omega, V, \alpha\}_{T, \Omega, P, \alpha}},
	\\
  & C_{\Omega, P, \Psi}=T \left(\frac{\partial S}{\partial T}\right)_{\Omega,
  P, \Psi}=T \frac{\{S, \Omega, P, \Psi\}_{T, \Omega, P, \alpha}}{\{T, \Omega,
P, \Psi\}_{T, \Omega, P, \alpha}},
	\\
  & C_{\Omega, V ,\Psi}=T \left(\frac{\partial S}{\partial T}\right)_{\Omega,
  V, \Psi}=T \frac{\{S, \Omega, V, \Psi\}_{T, \Omega, P, \alpha}}{\{T, \Omega,
V, \Psi\}_{T, \Omega, P, \alpha}}.
\end{align}
For example, the explicit calculation for $C_{J, P,\alpha}$ in $(T, \Omega,
P,\alpha)$ equilibrium space looks like
\begin{equation}\label{key}
  C_{J, P,\alpha}=T \left(\frac{\partial S}{\partial T}\right)_{\!\! J,P,\alpha}\!=T
  \frac{\{S, J, P, \alpha\}_{T, \Omega, P, \alpha}}{\{T, J, P, \alpha\}_{T,
    \Omega, P, \alpha}}=T\frac{\left|
		\begin{array}{cccc}
    \frac{\partial S}{\partial T} & \frac{\partial S}{\partial \Omega }&
    \frac{\partial S}{\partial P} & \frac{\partial S}{\partial \alpha } \\[5pt]
    \frac{\partial J}{\partial T} & \frac{\partial J}{\partial \Omega
    } & \frac{\partial J}{\partial P} & \frac{\partial J}{\partial \alpha
  } \\[5pt]
    \frac{\partial P}{\partial T} & \frac{\partial P}{\partial \Omega
    } & \frac{\partial P}{\partial P} & \frac{\partial P}{\partial \alpha
  } \\[5pt]
    \frac{\partial \alpha }{\partial T} & \frac{\partial \alpha }{\partial
    \Omega } & \frac{\partial \alpha }{\partial P} & \frac{\partial \alpha
  }{\partial \alpha }\\[5pt]
		\end{array}
		\right|}{\left|
		\begin{array}{cccc}
		\\[-10pt]
    \frac{\partial T}{\partial T} & \frac{\partial T}{\partial \Omega }&
    \frac{\partial T}{\partial P} & \frac{\partial T}{\partial \alpha } \\[5pt]
    \frac{\partial J}{\partial T} & \frac{\partial J}{\partial \Omega
    } & \frac{\partial J}{\partial P} & \frac{\partial J}{\partial \alpha
  } \\[5pt]
    \frac{\partial P}{\partial T} & \frac{\partial P}{\partial \Omega
    } & \frac{\partial P}{\partial P} & \frac{\partial P}{\partial \alpha
  } \\[5pt]
    \frac{\partial \alpha }{\partial T} & \frac{\partial \alpha }{\partial
    \Omega } & \frac{\partial \alpha }{\partial P} & \frac{\partial \alpha
  }{\partial \alpha }\\[5pt]
		\end{array}
		\right|}=T\frac{\left|
		\begin{array}{cccc}
			\frac{\partial S}{\partial T} & \frac{\partial S}{\partial \Omega }&
			\frac{\partial S}{\partial P} & \frac{\partial S}{\partial \alpha } \\[5pt]
			\frac{\partial J}{\partial T} & \frac{\partial J}{\partial \Omega
			} & \frac{\partial J}{\partial P} & \frac{\partial J}{\partial \alpha
			} \\[5pt]
			0 & 0 & 1 & 0 \\[5pt]
			0 & 0 & 0 & 1\\[5pt]
		\end{array}
		\right|}{\left|
		\begin{array}{cccc}
			\\[-10pt]
			1 & 0 & 0 & 0 \\[5pt]
			\frac{\partial J}{\partial T} & \frac{\partial J}{\partial \Omega
			} & \frac{\partial J}{\partial P} & \frac{\partial J}{\partial \alpha
			} \\[5pt]
			0 & 0 & 1 & 0 \\[5pt]
			0 & 0 & 0 & 1\\[5pt]
		\end{array}
		\right|},
\end{equation}
where we note that all derivatives of our parameters $(T,\Omega,P,\alpha)$ are equal to zero or one.

The expressions for the specific heats from the list above follow from the Nambu bracket formalism, introduced by \cite{Mansoori:2014oia}.

%====================
%	Bibliography
%====================
\newpage
\bibliographystyle{utphys}
\bibliography{References}
\end{document}